\documentclass[prb,twocolumn,superscriptaddress,longbibliography,
aps,prx,floatfix,preprintnumbers]{revtex4-2}

\usepackage[normalem]{ulem}
\usepackage{graphicx}
\usepackage{bm}
\usepackage{color}
\usepackage{amsmath}
\usepackage{amssymb}
\usepackage{epstopdf}
\usepackage{lipsum}
\usepackage{gensymb}
\usepackage{threeparttable}
\usepackage{multirow}
\usepackage{scrextend}
\usepackage{xfrac}
\usepackage[american]{babel}

\usepackage[urlcolor=blue,colorlinks=true,citecolor=blue,linkcolor=blue,pdfstartview={FitH},bookmarks=false]{hyperref}

\usepackage[dvipsnames]{xcolor}
\definecolor{mygreen}{rgb}{0.0, 0.6, 0.0}
\definecolor{pjorange}{rgb}{0.8, 0.3, 0.0}
\definecolor{jlblue}{rgb}{0.2, 0.5, 0.7}

\graphicspath{{fig/}{./fig/}{.}}

\begin{document}

\title{Contrasting anisotropic electron-phonon-spin coupling in Fe$_{3}$GeTe$_{2}$ and Fe$_{5}$GeTe$_{2}$: A helicity-resolved Raman study}

\author {Smrutiranjan Mekap}
\affiliation {Department  of  Physics, Indian Institute of Technology Kharagpur, Kharagpur 721302, India} 

\author {Jyoti Saini}
\affiliation{School of Physical Sciences, Jawaharlal Nehru University, New Delhi-110067, India}

\author{Andrzej~Ptok}
\affiliation{Institute of Nuclear Physics, Polish Academy of Sciences, ul. W. E. Radzikowskiego 152, 31-342 Krak\'{o}w, Poland}

\author{Pawan Kumar Srivastava}
\affiliation{School of Mechanical Engineering, Sungkyunkwan University, Suwon 16419, Republic of Korea}

\author{Changgu Lee}
\affiliation{School of Mechanical Engineering, Sungkyunkwan University, Suwon 16419, Republic of Korea}
\affiliation{SKKU Advanced Institute of Nanotechnology (SAINT), Sungkyunkwan University, Suwon 16419, Republic of Korea}

\author {Subhasis Ghosh}\email{subhasis.ghosh.jnu@gmail.com}
\affiliation{School of Physical Sciences, Jawaharlal Nehru University, New Delhi-110067, India}
\author {Anushree Roy}
\email{anushree@phy.iitkgp.ac.in}
\affiliation {Department  of  Physics, Indian Institute of Technology Kharagpur, Kharagpur 721302, India}
	\date{\today}
 
\begin{abstract}

Two-dimensional van der Waals ferromagnets Fe$_3$GeTe$ _2$ (F3GT) and Fe$_5$GeTe$_2$ (F5GT) exhibit pronounced magneto-optical responses, which open promising platforms for investigating the interplay among lattice, electronic, and magnetic degrees of freedom. Here, we present a comparative study of optical resonance-induced anisotropic electron-phonon coupling and its association with magnetic ordering in these systems using wavelength- and temperature-dependent helicity-resolved Raman spectroscopy. By resolving the doubly degenerate E modes under left- and right-circularly polarized excitations, we demonstrate that the temperature evolution of the chiral mode splitting ($\Delta f$) does not track the magnetization behavior, indicating that the helicity-dependent Raman response arises not solely from time-reversal symmetry breaking due to magnetic order, but also from spin-orbit-coupled electronic interactions. 
Notably, in F3GT, the out-of-plane magnetization indirectly governs the in-plane anisotropic electron-phonon coupling under optical resonance, whereas F5GT exhibits static anisotropic interactions. 
The Fano asymmetry parameter $1/q$ reveals mode- and temperature-dependent coupling strengths between phonons and the electronic continuum, with pronounced angular anisotropy in F3GT but isotropic behavior in F5GT--- a consequence of its multiple Fe sites and enhanced interlayer hybridization in the latter. 
Our results demonstrate the role of crystal structure and magnetic anisotropy in shaping the anisotropically coupled electron-phonon-spin dynamics in these layered metallic ferromagnets,  
and highlight Fe$_n$GeTe$_2$ as a versatile platform for microscopic insight into chiral light-matter interactions in layered metallic ferromagnets.

\end{abstract}

\maketitle

\section{introduction}
\hspace*{0.5cm}
The emergence of intricate spin dynamics in two-dimensional van der Waals (2D vdW) magnets has drawn special attention from a physics perspective by demonstrating the anomalous topological Hall effect~\cite{miao2023,casas2023}, ultrafast demagnetization~\cite{lichtenberg2022}, topological spin textures such as magnetic bubbles~\cite{lv2024,khela2023}, spin reorientation~\cite{ly2021}, and the generation of spin waves~\cite{zhang2020a,zhang2020b,afanasiev2021,schulz2023}. Interest in these systems also arises from the viewpoint of their applications in 2D spintronic device fabrication and all-optical magnetic switching~\cite{adhikari2025,dabrowski2022,yang2021}. Among various 2D magnetic materials, iron-based compounds with the formula Fe$_n$GeTe$_2$ (where $n$=3--5) have attracted particular attention due to their remarkable magnetic and structural properties~\cite{liu2022,ren2023,adhikari2025,seo2020,jiang2022}, which result from the existence of non-equivalent Fe atoms in their unit cell. Particular interest has grown in Fe$_3$GeTe$_2$ (F3GT)  and Fe$_5$GeTe$_2$ (F5GT) because of their elevated Curie temperature ($T_C$)~\cite{chen2013,may2016,may2019a,li2018,zhang2020c}, a necessary criterion to achieve robust magnetic
ordering at room temperature for any application. The value of $T_C$ is $\sim$220 K for F3GT~\cite{may2016,chen2013} and $\sim$ 275 K  for F5GT~\cite{chen2022,chen2023,may2019a,may2019b,zhang2020c}.
F3GT and F5GT are composed of Fe-rich slabs sandwiched between Te layers. The number of Fe sublayers plays a key role in strengthening exchange interactions and stabilizing magnetic ordering in these systems~\cite{liu2022,jiang2022}.  The weak interlayer vdW bonding leads to pronounced magnetic anisotropy, with stronger in-plane than out-of-plane spin correlations, or vice versa, and supports the emergence of competing magnetic interactions and complex spin textures under applied fields~\cite{Bao2022,tan2018}.

For such an anisotropic 2D magnetic lattice, phonons, which correspond to the rotation of atomic displacements, also referred to as chiral phonons, explain spin-phonon  ~\cite{du2019} and spin-orbit-exciton interactions ~\cite{lujan2024,kong2024}, as well as electron-phonon interactions~\cite{chen2018,Zhu2025}. These phonons carry well-defined pseudoangular momentum (PAM), which can be probed using circularly polarized light in Raman scattering~\cite{lyu2020,yin2021} and circular dichroism~\cite{chen2018},  with photons carrying spin angular
momentum. The PAM associated with chiral phonons arises from the breaking of structural or time-reversal symmetry (TRS), caused by external strain~\cite{rostami2022}, a magnetic field~\cite{chen2018}, or intrinsic magnetic anisotropy~\cite{che2025}.

In the past decade, significant progress has been made in establishing the efficacy of helicity-resolved Raman spectroscopy to study magnetism in 2D magnetic systems, like VI$_3$~\cite{lyu2020}, CrBr$_3$~\cite{yin2021}, CoSn-like compounds~\cite{che2025}, topological magnetic insulators~\cite{chen2025,kobialka2022}, and others~\cite{sun2021}. The symmetry of the Raman tensor changes under broken TRS. The differential sensitivity of the phonon modes of a given symmetry near the $\Gamma$ point to left and right circularly polarized light reflects TRS breaking, which is concomitant with spin-flip transitions in the 2D vdW magnets~\cite{lyu2020,chen2023}. The literature is especially rich in demonstrating helicity conservation and helicity reversal of the phonon modes of different symmetries in anisotropic materials due to complex Raman tensor elements arising from strong electron-phonon interactions, particularly involving valley-polarized electrons of 2D materials, such as graphene, MoS$_2$, WSe$_2$, etc., under an externally applied magnetic field~\cite{han2022, Zhang2015,chen2018,liu2022}. In the case of the quantum magnet CoTiO$_3$~\cite{lujan2024,Prosnikov2026}, the effect on chiral phonon modes due to their strong hybridization with spin-orbit excitons could be observed in magneto-Raman measurements, suggesting the possibility of chiral phonons to couple with topological bosons in the case of energy matching~\cite{lujan2024}. Raman spectroscopy has been used to demonstrate the existence of helical phonon modes in F3GT~\cite{du2019}.

The contrasting symmetry of the crystal structure, magnetic anisotropy, spin dynamics, and electronic structures of F3GT and F5GT make them ideal systems for a comparative helicity-resolved Raman study to elucidate the role of the above mentioned parameters in determining magneto-optical effects in metallic van der Waals ferromagnets.
As mentioned above, while chiral phonons have been widely studied in semiconducting nonmagnetic materials, their interplay with magnetic ordering and spin-orbit-coupled electronic states in metallic ferromagnetic vdW systems remains poorly understood. Here, we present a comparative analysis of phonon dynamics and its coupling with spin and electronic degrees of freedom in both compounds using helicity-resolved Raman spectroscopy under co- and cross-polarization configurations. The observed splitting of the chiral modes under left- and right-circularly polarized (LCP and RCP, respectively) excitations, together with its distinct temperature evolution compared to the magnetization, provides an optical signature of TRS breaking associated with the magnetic order and further suggests an important role of spin-orbit-coupled electronic interactions.  The polar plot of the scattered intensity under RCP and LCP incident excitation reveals distinct electron-phonon coupling characteristics in F3GT and F5GT. The Fano spectral line shape, in addition to the angular and temperature dependence of the Fano parameters for both systems, are analyzed to establish the same. The order-parameter-like co- and cross-circular Raman asymmetry parameters, which serve as sensitive probes of helicity-dependent magneto-optical activity and symmetry-selective spin--orbit-coupled interactions, respectively, are compared for both systems. The contrasting behavior of F3GT and F5GT elucidates the role of magnetic anisotropy and crystal structure in engineering helicity-selective spin–phonon interactions in layered magnetic materials.

 \section{Experimental details}

The single crystal  F3GT and F5GT samples were grown using the chemical vapor transport (CVT) method; details of the sample growth are available in Ref.~\cite{Saini2026}.
The detailed characteristics, crystal dimensions, and x-ray diffraction (XRD) patterns of the present F3GT and F5GT samples used by us are published elsewhere \cite{Saini2026}. Temperature-dependent magnetic measurements in field cooled (FC) mode were carried out in the physical properties measurement system (PPMS) by applying a magnetic field of $\mu_0H=0.01$ T as the crystal was cooled from room temperature to a low temperature. Measurements were carried out with a magnetic field applied parallel to the $ab$-plane and the $c$-axis. The intriguing and complex magnetic behaviors of these compounds are experimentally established \cite{Tiwari2024,Saini2026}. 

Micro-Raman measurements were performed in back-scattering geometry using a 50$\times$L (long-working-distance objective) lens in a triple-stage monochromator Raman spectrometer T64000 (Horiba, France). The spectrometer is equipped with a confocal microscope (Olympus, Japan) and a Peltier-cooled charge-coupled device (Syncerity USA). A 532 nm Nd-YAG laser was used as the excitation source.  For wavelength-dependent Raman measurements, a multiwavelength Ar$^+$-Kr$^+$ gas laser (Innova 70C, Coherent, USA) was used.  Spectra were recorded at different laser powers to check the heating effect by monitoring the shifts in the Raman mode. Following this, all reported spectra were recorded with a laser power of 1 mW on the sample. Temperature-dependent Raman spectroscopy was carried out using a low-temperature
sample stage THMS600 (Linkam, USA), equipped with a cryostat and liquid nitrogen pump.  
For helicity-resolved measurements, the laser was guided by a polarizer and a half-wave plate (HWP). A quarter-wave plate (QWP) was placed at the common path of the incident and scattered paths to convert linear to circular polarization. Circularly polarized scattered light was detected after it passed through an HWP and an analyzer. For angle-resolved measurements, the HWP in the scattered light path was rotated at regular angular intervals to select different helicities of the scattered light. The schematic diagram of the setup is shown in Fig. S2(b) of Ref.~\cite{mekap2026}. Helicity-resolved Raman measurements were carried out in IJ (I, J= RCP/LCP) configurations,
where I and J represent the polarization states of the incident and scattered light, respectively. In these measurements, the projection of the J-component of the scattered light was recorded for incident light with I state. For example, the RL configuration corresponds to RCP incident light and the detection of the LCP component of the scattered radiation.

\section{Crystal structure and magnetic anisotropy}

\begin{figure}[h!]
\centering
	\includegraphics[width=\linewidth]{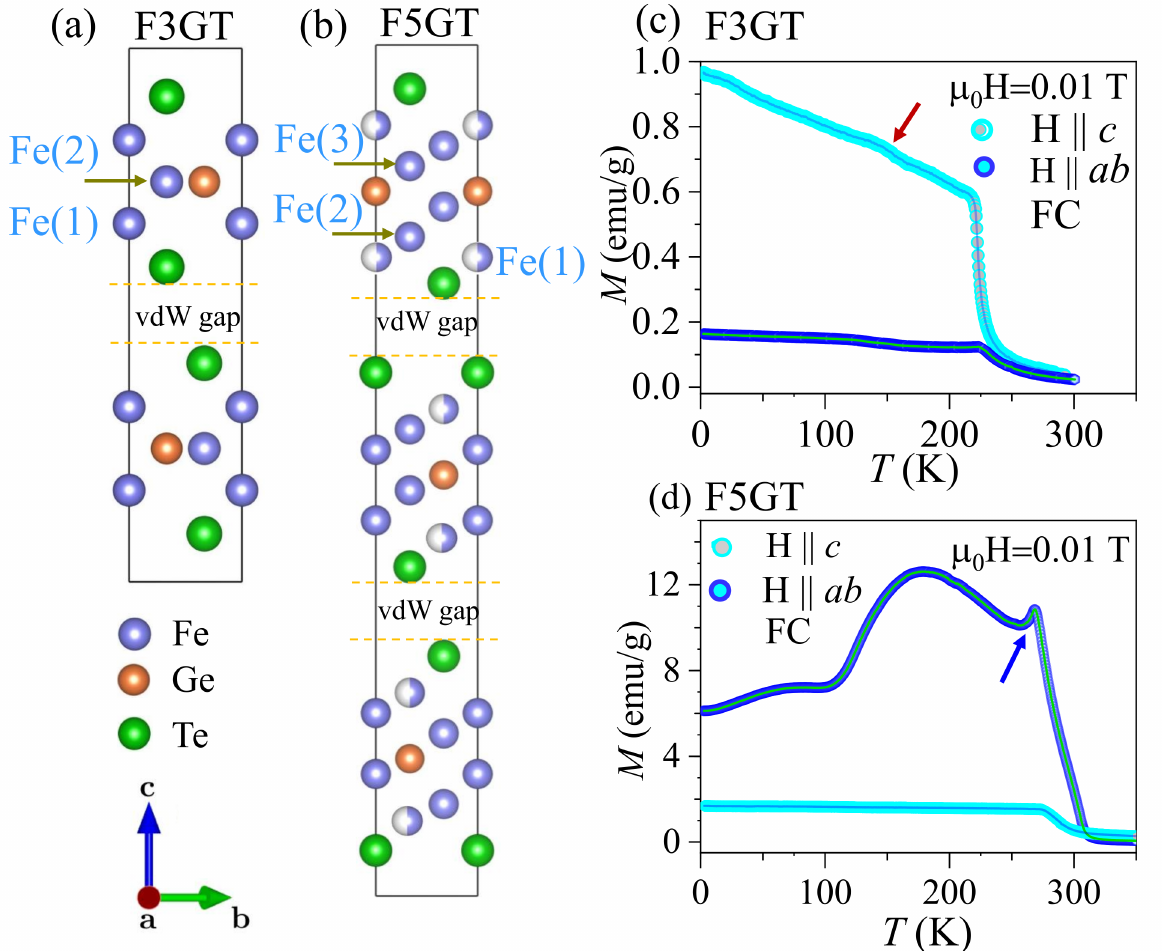}
	\caption{Crystal structure of (a) F3GT, and (b)  F5GT. Fe, Ge, and Te atoms are shown with colored solid balls. vdW gaps between layers are shown by a pair of yellow dashed lines. The polarised field-cooled (FC) $M$--$T$ curves measured for in-plane and out-of-plane magnetic fields for (c) F3GT, and (d) F5GT. For descriptions of arrows, refer to the text.}
	\label{Crystal structure}
\end{figure}

Unit cells in the crystal structures for F3GT and F5GT are shown in Figure \ref{Crystal structure}(a) and (b), respectively. F3GT crystallizes in a hexagonal structure with  $P6_{3}/mmc$ symmetry (space group no. 194), while F5GT adopts a rhombohedral structure belonging to $R3m$ symmetry (space group no. 160). Single-crystal XRD patterns of both compounds are reported elsewhere~\cite{Saini2026}. As mentioned earlier, both systems consist of layered Fe/Ge atoms in Fe/Ge slabs sandwiched between Te layers. The unit cell of F3GT consists of two monolayers, where each monolayer consists of three Fe atoms, one Ge atom, and two Te atoms. In contrast, the unit cell of F5GT comprises three monolayers; each monolayer contains six atomic layers, which fundamentally distinguishes its structural and magnetic properties from those of F3GT.  F3GT hosts two inequivalent Fe sites (Fe(1) and Fe(2)). In F5GT, the layer symmetry
leads to three non-equivalent Fe positions, while Fe(1) has a refined fractional occupancy ($0.5$)~\cite{may2019a}.  This leads to increased structural disorder and additional lattice distortions, including larger vdW gaps in F5GT than in F3GT. The increased Fe-layer and multiple Fe oxidation states in F5GT lead to complex magnetic interactions and electronic correlations in the system.

Figure~\ref{Crystal structure}(c) and (d)  present the temperature-dependent magnetization measured with a magnetic field of $\mu_0H=0.01$ T, applied along the $ab$- plane and the $c$- axis for F3GT and F5GT, respectively. Both compounds undergo a paramagnetic (PM)-to-ferromagnetic (FM) transition, with $T_C \approx 223$ K for F3GT and a higher $T_C \approx 273$ K for F5GT, reflecting the enhanced exchange interactions arising from additional Fe layers in the latter. Both compounds exhibit strong magnetic anisotropy, with in-plane magnetization exceeding out-of-plane magnetization for F5GT, whereas the reverse is true for F3GT. The anisotropy is stronger in F3GT than in F5GT. 
In F3GT, the kink at $\approx$ 160 K  (shown by a red arrow)  in the magnetization plot originates from the competing FM and antiferromagnetic (AFM) interactions, consistent with reports of field-dependent interlayer magnetic ordering and strong out-of-plane anisotropy~\cite{Yi2017}. In contrast, F5GT displays more complex magnetic behavior, with multiple magnetic anomalies associated with competing FM-AFM interactions along with noncollinear spin-spin interactions. In F5GT, the kink at 273 K (shown by a blue arrow) has been attributed to a magnetic anomaly~\cite{zhang2020c}. 

Both systems exhibit itinerant magnetism originating from Fe-derived electronic states. Due to the presence of multiple Fe sites, F5GT exhibits higher electronic correlations and magnetic complexity than F3GT. These contrasting magnetic anisotropies and ordering phenomena provide a direct framework for understanding the observed differences in phonon splitting, Fano asymmetry, and helicity-dependent Raman responses in F3GT and F5GT.

 \section{Results and discussion}

\subsection{Helicity resolved Raman measurements:}

The conservation of angular momentum in a Raman scattering process involving circularly polarized excitation demands $\vec{L}_{in}=\vec{L}_{out}+\vec{L}_{ph}+\vec{L}_{lattice}$~\cite{Zhang2015,kim2025}. 
$\vec{L}_{in}$ and $\vec{L}_{out}$ are the angular momenta of the incident light and the scattered light, respectively. $\vec{L}_{ph}$ of the chiral phonons is described as PAM, since rather than continuous rotational invariance, the origin is its discrete rotational symmetry. $\vec{L}_{lattice}$ 
 represents the angular momentum associated with the crystal background, excluding the explicitly identified chiral phonon mode.
Correspondingly, the conservation of the projected angular momentum along a chosen quantization axis  (say $z$)  can be written as $l_{in}^{z}=l_{out}^{z}+l_{phonon}^{z}+l_{lattice}^{z}$.
For $n$-fold rotational crystal symmetry, the PAM of the phonon of wavevector $\vec{q}$, $l_{ph,q}$,  can be defined by the following
$\hat C_{n}u_{q}=e^{(-2\pi i/n)l_{ph,q}}u_q$, where $\hat C_n$ is the $n$-fold rotation operator, $u_q$ is the phonon Bloch wave function ~\cite{Zhang2015,zhang2022}. The chiral phonons have a nonzero projection of PAM $\sim$ $\pm\hbar$. As $l_{lattice}$ is absorbed by
rigid crystal rotation~\cite{Zhang2015,kim2025}, the conservation relation for Raman scattering in backscattered geometry is simply expressed as $l_{in}=l_{out}+l_{ph}$.
 
\begin{figure}[h!]
\centering
	\includegraphics[width=0.75\linewidth]{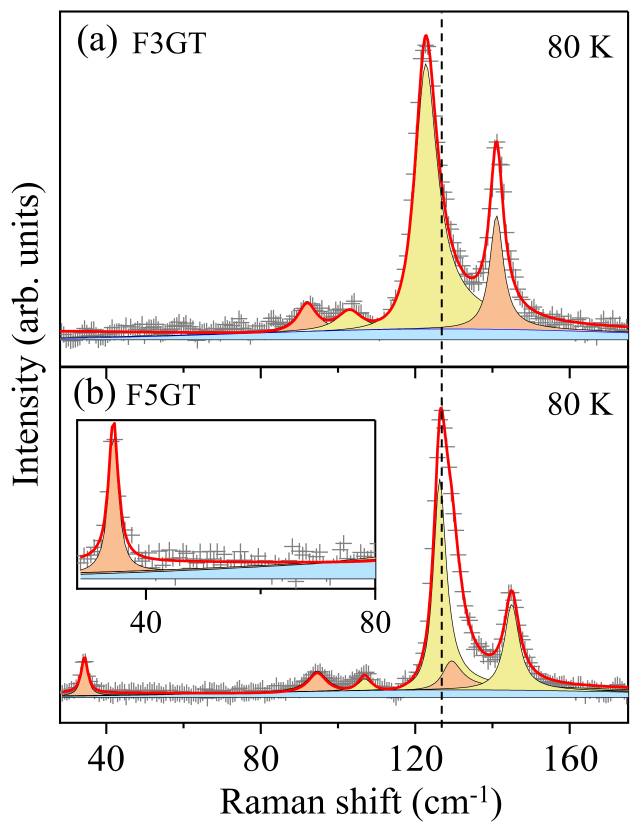}
	\caption{Raman spectra of (a) F3GT, and (b) F5GT at 80 K under RL  configuration. The measured spectra are shown with  + symbols, and the net fitted spectra are shown by red curves. The deconvoluted A and E modes are shown by orange and yellow shaded areas, respectively. The black dashed line marks the difference in Raman shift of these two compounds for an A$_1$ mode. The broad background is shown by blue shaded area. For magnified view of the background, refer to middle panels of Fig. S8. The inset of (b) shows the magnified spectral range
 between 25 and 80 cm$^{-1}$.}
	\label{spectra}
\end{figure}
 In F3GT and F5GT, chiral phonon modes arise due to the six-fold and three-fold
rotational symmetries, respectively, of the systems. Figure~\ref{spectra} plots Raman spectra recorded at RL polarization configuration for (a) F3GT and (b) F5GT.
Because of chemical similarity, though F3GT and F5GT exhibit very similar overall Raman responses, except a small shift between the main peaks of these two compounds (shown by red dashed lines), details of the spectral profile and how they evolve across magnetic transitions, which we discuss in this article, are strikingly different, revealing how Fe stoichiometry and lattice complexity tune electron--phonon--spin interactions.
The irreducible representation of the optical phonon modes of the P6$_{3}/mmc$ (194) space group of F3GT is $\Gamma$= 2A$_{1g}$+2E$_{1g}$+4E$_{2g}$~\cite{Milosavljevi2019}. In Figure~\ref{spectra}(a), we identified four Raman-active modes over the spectral range between 25 and 350 cm$^{-1}$ in F3GT in RL configuration. 
Similarly, for F5GT, with space group $R3m$ and corresponding irreducible representation for the optic Raman mode as $\Gamma$= 7A$_{1}$+ 7E, we observed 6 Raman peaks shown in Figure~\ref{spectra}(b) over the spectral range between 25 and 350 cm$^{-1}$ in circularly-polarized configurations. A higher number of modes, concomitant with the expected Raman shift for F3GT and F5GT, could be identified in the linearly polarized configuration [see Sec.~S1 in the  Supplemental
Material (SM)  \cite{supplementary}]. 

Here we would like to mention that the spectral profiles of both F3GT and F5GT vary considerably across the literature \cite{du2019,Milosavljevi2019,xu2020b,cai2023,liang2023}. Fe vacancies \cite{Milosavljevi2019,yadav2024}, layer thickness \cite{weerahennedige2024,kong2021}, surface modification \cite{du2019}, and tellurium precipitation \cite{Antony2025} are the main reasons for such a contradictory spectral nature. 
The  XRD pattern \cite{mekap2026,Saini2026} of these compounds does not exhibit any evidence of Fe-vacancy-induced structural modification. To verify the intrinsic origin of the Raman spectra of F3GT and F5GT, and to exclude a dominant contribution from TeO$_2$ or elemental Te, we performed a series of additional control measurements, see Sec. S2 in SM \cite{supplementary}. Unpolarized Raman spectra of TeO$_2$ and elemental Te were measured under the same experimental conditions and compared with those of F3GT and F5GT. The reported TeO$_2$ and Te spectra also consistent with literature \cite{Smirnov2018,Manjón2021}. Although Te-related vibrations occur in a similar frequency range, the characteristic peak positions, the separation between two peaks under discussion, and overall spectral profiles of TeO$_2$ and elemental Te are distinct from those observed in the two FGT compounds. In particular, the additional characteristic features expected for TeO$_2$  and elemental Te are absent in the FGT spectra.  Furthermore, spectra acquired from freshly cleaved and uncleaved FGT crystals exhibit essentially identical Raman profiles, whereas measurements at different excitation wavelengths reveal distinctly different trends in phonon mode intensities \cite{NAVEEDULHAQ2024}. In case of F3GT,  the spectra are also consistent with previously reported Raman responses of bulk F3GT ~\cite{weerahennedige2024}. Together, these observations strongly support an intrinsic origin of the observed Raman modes and disfavor a dominant contribution from surface TeO$_2$ or elemental Te. Further details of these control measurements are provided in Sec. S2 in SM \cite{supplementary}.

Raman spectra of F3GT and F5GT exhibit a broad background extending below the main phonon features in the range of 25–175 cm$^{-1}$. Given the itinerant nature of the electronic states in these compounds, the observed continuum is consistent with an electronic Raman background that can interfere with the discrete phonon excitations, thereby providing a basis for describing the phonon line shapes using a Fano profile. Further details, including the temperature- and excitation-energy dependence of the continuum, are presented in the Sec. S3 in SM \cite{supplementary}. Furthermore,
in the literature, independent evidences suggest that phonon–continuum interactions are relevant in FGT. Fano-type behavior has previously been reported in FGT using techniques other than Raman spectroscopy. For example, Zhang et al. in Ref. \cite{zhang2018} reported Fano-type features in STM measurements of F3GT. In that case, the interference arises from different tunneling pathways and therefore is not identical to the Raman process considered here. More recently, Chauhan et al. \cite{chauhan2026} reported Fano interference between a coherent optical phonon and an electronic continuum in F3GT using THz pump-probe measurements and discussed the temperature evolution in terms of the overlap of the phonon response with a dissipative electronic continuum. These results provide independent evidence that coupling between phononic excitations and electronic continua is physically relevant in FGT.

  We fitted the observed spectra using a phenomenological Gaussian background $B_{G}(\omega)$ (see  Fig. S7 in Sec. S3 \cite{supplementary}) together with either a Fano or Lorentzian phonon profile.
\begin{eqnarray}
\nonumber I ( \omega ) = B_G(\omega)+ \sum_{i} \frac{A_i}{2\pi} \frac{ \Gamma_{i} }{ (\omega - \omega_{i})^{2} + (\Gamma_{i}/2)^{2} }\\ + \sum_{j}\frac{F_{j} (q_{j}+\epsilon_{j})^2 }{ (1+\epsilon_{j}^2) }.
\label{net intensity} 
\end{eqnarray}

In the above relation, the second term corresponds Lorentzian spectral profile of all Raman modes, with Raman shift, width, and integral intensity as $\omega,\Gamma$ and  $A_{i}$, respectively, except for the prominent modes at 122 cm$^{-1}$ and 142 cm$^{-1}$ for F3GT. These two modes are fitted with a Fano line shape, given by the third term in Eq.~(\ref{net intensity}). $F$ is a constant, and $\epsilon_{j}=\frac{(\omega-\omega_{0j})}{\Gamma_j}$, along with $\omega_{0j}$ and $\Gamma_{j}$, represents the peak position and width, while $q$ is the asymmetry parameter. Using angle-resolved intensity variation of these modes under linearly polarized incident radiation (see Sec.~S4 in the SM \cite{supplementary}, and Ref.~\cite{Milosavljevi2019}), we ascribe A$_1$ and E symmetries to the observed Raman modes at 125 cm$^{-1}$ and 142 cm$^{-1}$.

For F5GT, Raman mode analysis from first-principles density functional theory (DFT) calculations~\cite{mekap2026} suggests the presence of three modes over the spectral window between 125 and 170 cm$^{-1}$. In light of this, we fitted this spectral range using three peaks with Fano line shapes. The prominent Raman peak at 35 cm$^{-1}$ is also fitted with a Fano line shape. The remaining observed modes are fitted with Lorentzian line shapes. Following the angle-resolved intensity variation of the prominent Raman modes (as shown in Sec.~S5(b) in the SM of Ref.~\cite{mekap2026}), and the expectation from DFT calculations, we assign the peaks at 125 and 143 cm$^{-1}$ to the A$_1$(1) and A$_1$(2) modes. The peaks at 35 cm$^{-1}$ and 129 cm$^{-1}$ to E(1) and E(2) symmetries. Because of the strong overlap of A$_1$(1) and E(2) modes, we will refrain from discussing the characteristics of their spectral parameters in this article. Our focus will be on E(1) and A$_1$(2) modes only. Due to the positive Fano parameter for the A$_1$(2) mode, i.e. asymmetry of its spectral wing towards higher wavenumber, its spectral features are affected minimally by the two nearby low-wavenumber Raman modes (A$_1$(1) and E(2)).  
In Figure~\ref{spectra}, the best-fitted curves to the data points are shown in red. In all panels of Figure~\ref{spectra}, the deconvoluted A$_1$ and E modes are shown by orange  and yellow shades, respectively.
\subsection{Lifting of degeneracy of chiral phonons}

\begin{figure*}[htbp]
\centering
	\includegraphics[width=0.9\linewidth]{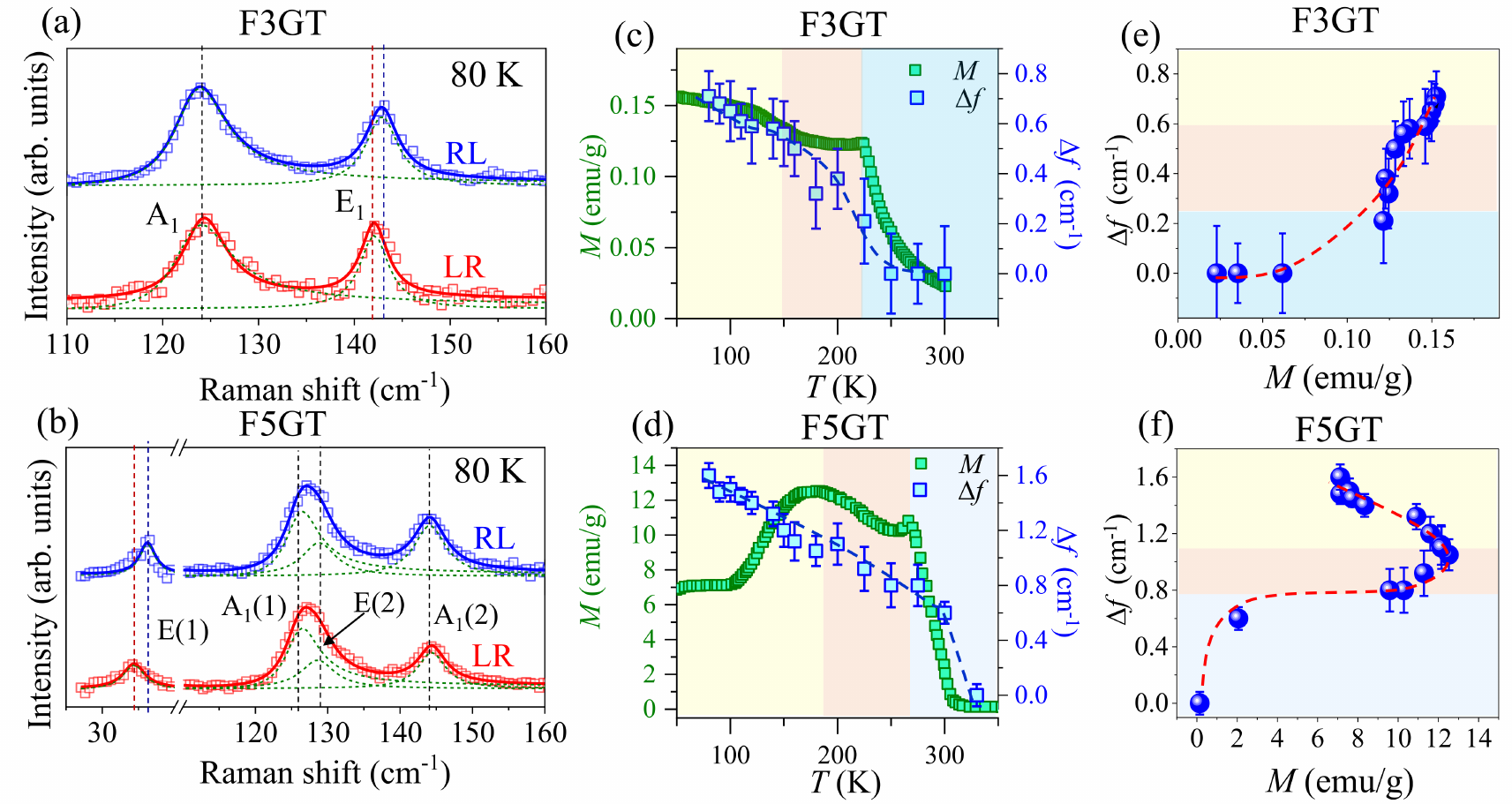}
	\caption{Magnified helicity-resolved fitted-background-subtracted Raman spectra in RL and LR configurations at 80 K are presented by blue and red symbols, respectively, for (a) F3GT, and (b) F5GT. The dashed lines mark the maximum of the Raman peak positions in these configurations. The comparison between $M-T$  in the $ab$-plane and $\Delta f-T$  for (c) F3GT  and (d) F5GT. The variation of $\Delta f$ with in-plane magnetization for (e) F3GT, and (f) F5GT.  Error bars with $\Delta f$ in panels (c)--(f) are estimated using the mean and standard deviation of the Raman shift obtained from a non-linear curve fitting procedure, and then following the propagation of error method. The color shades mark the one-to-one correspondence between $\Delta f$ and $M$ across three temperature ranges (blue, orange, and yellow) characterized by different spin-correlations. The blue areas correspond to paramagnetic phase, yellow the FM ground state, and the orange shaded areas mark a competing FM and AFM phases in F3GT \cite{Yi2017} and the anomalous temperature range due to non-unidirectional magnetic moments between the monolayers in F5GT \cite{ly2021,zhang2020c}.}
	\label{RL_LR}
\end{figure*}

The splitting of chiral E phonon modes is a consequence of TRS breaking or, sometimes, structural chirality. Osihi {\it et al.}~\cite{oishi2024} reported the selectivity of chiral phonons in quartz to polarization configurations. They observed a slight difference in Raman shift in E modes when probed with cross-circular polarization configurations and attributed it to the consequence of the splitting of $\Gamma$ point phonon modes into 
two phonon branches, as under circularly polarized light, the degenerate chiral E modes carry opposite angular momenta. In the case of CoTiO$_3$, the splitting of the E mode is observed under an external magnetic field due to the strong magneto-phononic effect originating from the spin-orbit excitation of phonons~\cite{lujan2024}.  Che {\it et al.}~\cite{che2025} observed the splitting of E modes in Weyl semimetal Co$_3$Sn$_2$S$_2$ without the application of a magnetic field below $T_C$. It is suggested that the chiral phonons can be controlled directly by the intrinsic magnetization of the compound rather than by an external magnetic field.

In both F3GT and F5GT, the doubly degenerate E modes consist of two components (E=E$^{+}\oplus$ E$^{-}$)  with opposite angular momentum ($\pm \hbar$). In our experimental configuration, i.e., in the backscattering geometry with light incident along the $c$-axis, RCP and LCP excitations selectively couple to the E$^{+}$ and E$^{-}$ chiral phonons carrying angular momentum $+\hbar$ and $-\hbar$, respectively. We define the difference in Raman shift in RL and LR configurations as  $\Delta{f}=f_{RL}-f_{LR}$, $f_{RL}$ and $f_{LR}$ are the peak positions in the respective configurations.
The magnified views of the Raman spectrum of the E and A$_1$ modes of F3GT in RL and LR polarization configurations are shown in Figure~\ref{RL_LR}(a). The same for F5GT at 80 K is shown in Figure~\ref{RL_LR}(b).  In the case of F3GT, the E mode shows a difference of $\Delta f = 0.7$~cm$^ {-1}$ between the RL and LR configurations. On the other hand, the E(1) mode of F5GT exhibits a relatively large Raman shift difference of $\Delta f = 1.6$~cm$^ {-1}$.

Characteristic Raman spectra of F3GT and F5GT, recorded in LR and RL configurations at various temperatures, are available in Fig. S12 in SM \cite{supplementary}. Figure~\ref{RL_LR}(c) and (d) plot the comparison between the evolution of $\Delta{f}$ and magnetization $M$ with  $T$ for F3GT and F5GT. 
Hence, Figure~\ref{RL_LR}(e) and (f) demonstrate the correlation between $\Delta f$ and $M$ for F3GT and F5GT, respectively. It is reasonable to assume that anharmonicity in vibrational potential or lattice expansion affects the phonon renormalization with temperature of degenerate E modes, E$^+$ and E$^-$, nearly equally; thus, it cancels out in the value of $\Delta f$. Assuming mean-field theory is still valid in these systems, the shift in phonon frequency due to spin-phonon coupling is expected to be proportional to $|M(T)|^2$ \cite{Granado1999}.  However, it is to be noted that phonon self-energy renormalization arising from spin-phonon coupling is a scalar quantity that is even under time reversal. 
Thus, the helicity-dependent splitting $\Delta f$ at a temperature $T$, originating from TRS breaking and, by symmetry, is expected to scale primarily with the TRS-breaking order parameter; hence,  approximately proportional to $M$.  Thus, the observed non-linear variation of $\Delta f (T)$ with $M(T)$ suggest that the chiral phonons are also sensitive to local spin correlations, spin fluctuations, and itinerant magnetic effects beyond the macroscopic magnetization. Earlier studies in the literature using DFT calculations reveal strong spin-orbit coupling (SOC) in F3GT~\cite{jiang2025} and F5GT~\cite{bera2024}. It has been shown that SOC-induced energy differences influence higher-order exchange interactions, which, in turn, affect the stability and formation of magnetic characteristics in F3GT~\cite{jiang2025}. Using DFT calculations, Bera {\it et al}~\cite{bera2024}. estimated an SOC-related anisotropy energy of about 6 meV in F5GT, indicating a significant contribution of the same to the magnetic behavior of these systems. 
Thus, we attribute the observed mismatch between $\Delta f$ and $M$ in Figure~\ref{RL_LR}(c)-(d) and non-linear variation of $\Delta f$ with $M$ to the above factors-driven phonon response in F3GT and F5GT.  
Here, we would like to mention that similar observations due to the coupling of spin, orbital, and lattice degrees of freedom have been attributed to the origin of the chiral response of phonons in the presence of an external magnetic field in  CoTiO$_3$~\cite{lujan2024}.

In helicity-resolved Raman measurements, one probes $\langle\vec {S}_{i}\times \vec {S}_{j}\rangle$, $\langle \vec{L}_{phonon}\cdot \vec{S}\rangle$, and TRS breaking. For complex magnetic structures, like in F5GT (in which, in addition to competing FM and AFM interactions, anisotropic interactions are present), they are not necessarily proportional.  Hence, we find a much more complex variation of $\Delta f$ with $M$ in case of F5GT than in F3GT. The larger value of $\Delta f$ in the case of F5GT than in F3GT (as shown in Figure \ref{RL_LR}(a)-(b)) can be attributed to the above factors. We also need to keep in mind that the chirality of the phonon modes strongly depends on the magnetic point group~\cite{zhang2026}, which is different for F3GT and F5GT. In Fig. \ref{RL_LR}(c)--(f), the light blue, light saffron, and light yellow shades mark PM phase, competing FM and AFM regime and FM  ground state, respectively, providing a visual guide to the evolution of the TRS breaking-induced splitting of doubly degenerate E modes across these distinct spin-correlation regimes.

Thus, helicity-resolved Raman measurements provide magnetic information well beyond conventional magnetic measurements. 

\subsection{Helicity-dependent electron–phonon response}

The existence of resonance-induced strong electron-phonon coupling in vdW 2D magnets has been reported using linearly polarized Raman measurements~\cite{mekap2026}. In F3GT and F5GT, the electronic states are spin-polarized.
Hence, the helicity-resolved Raman spectroscopy provides significant advantages over conventional linear-polarization measurements for probing strong electron-phonon coupling in F3GT and F5GT. By selectively exciting the two counter-rotating components of doubly degenerate E modes, helicity-resolved Raman measurements unravel the angular-momentum-dependent anisotropic coupling of phonons to spin-polarized electronic states via magneto-chiral electron-phonon interactions. This allows one to study the chirality-dependent Fano asymmetry and the correlation between the electron-phonon strength, magnetic order, and anisotropy. Thus, while linear-polarization Raman measurements reveal anisotropic electron-phonon coupling governed by crystal symmetry~\cite{mekap2026}, helicity-resolved Raman measurements can be considered as an essential tool for understanding the interplay between lattice dynamics, electronic structure, and magnetism in these systems.

\begin{figure*}[htbp]
\centering
	\includegraphics[width=\linewidth]{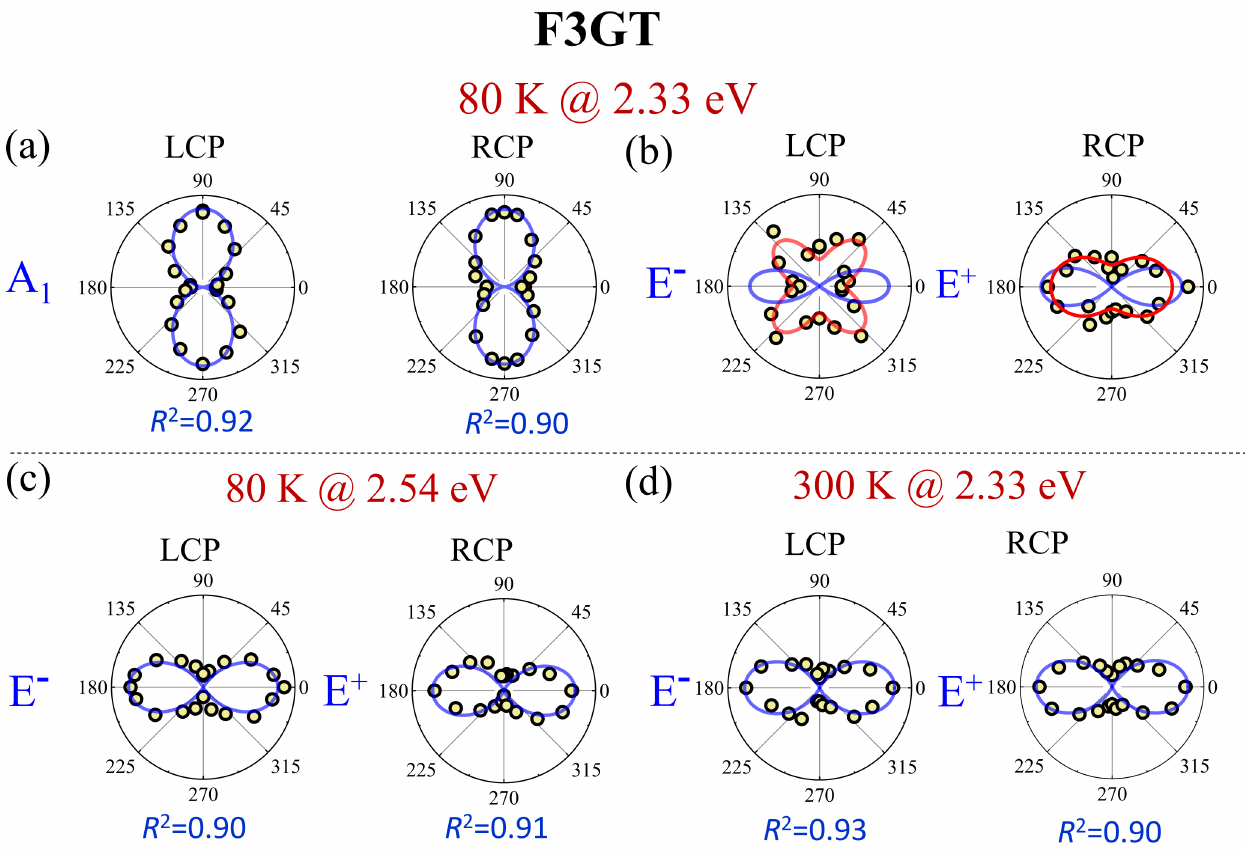}
	\caption{F3GT: Polar plots of normalized Raman intensity for in LCP and RCP configurations at 80 K of (a) the A$_1$ mode, and (b) of the E$_1$ mode using 2.33 eV as the excitation energy. (c) Polar plots of normalized Raman intensity of the  E$_1$ modes with 2.54 eV excitation energy at 80 K. (d) The same with 2.33 eV at 300 K. 
    In all polar plots, data points are represented by olive-colored solid balls.  Except for panel (b), in all panels, solid blue curves are the result of the best fit to the data points using the Raman tensor in Eq.~(\ref{Eqn_6m}).    The corresponding values of $R^2$ are available below each plots using blue font. In (b), the blue curves only show the expected ones, not the fit. Red curves guide the eyes along the measured data points.}
	\label{polar_circular_F3GT}
\end{figure*}

Figure~\ref{polar_circular_F3GT}(a)--(b) present the polar plots of the intensity of the A$_1$ and E$_1$ modes of F3GT at regular intervals of scattering directions for the incident RCP and LCP light at 80 K. 
In non-resonant helicity-resolved Raman scattering, the scattering intensity for the Raman tensor $\Re$ follows the relation $I(\varphi)\propto\left|\vec{P}_{\sigma s}^{T}\cdot \Re\cdot \vec{P}{\sigma}_i\right|^2$. $\vec{P}_{\sigma i}$ and $\vec{P}_{\sigma s}$ are the Jones vectors for the incident and scattered circularly polarized light. 
The helicity, $\sigma$, of a photon is defined as an eigenvalue of the matrix operator $\hat{S}_z$, $z$ component of the SO(3) vector rotation. 
For the incident RCP light along the $z$ ($c$-axis),
$\vec{P}_{\sigma i}^{R}
=\frac{1}{\sqrt{2}}$ 
$\begin{bmatrix} 1 & i & 0 \end{bmatrix}$,  $\vec{P}_{\sigma s}$ carries the projection of both co- and cross-polarized components for a given detection angle $\varphi$ and is given by $\vec{P}_{\sigma s}^{T}=\cos{\varphi}(L)+\sin{\varphi}(R)=\frac{1}{\sqrt{2}}$
$\begin{bmatrix} \cos{\varphi}+\sin{\varphi} & i(\cos{\varphi}-\sin{\varphi}) & 0 \end{bmatrix}$. A similar expression can be derived for LCP incident radiation.
For a magnetic crystal, the  Raman scattering matrices can be obtained by extending Loudon's theory \cite{Loudon01964} of Raman scattering for the irreducible corepresentations of the magnetic point groups of the corresponding space group. 
We consider the  $6/mmm$ magnetic point group for $P6_{3}/mmc$ space group for F3GT \cite{cracknell1969}. Further,  due to the breaking of TRS in the FM state, the antisymmetrical
imaginary elements are allowed in the Raman tensor (see Sec. S4 of SM \cite{supplementary}).
The Raman tensors for achiral A$_1$, and degenerate chiral E modes are,
\begin{equation}
\begin{aligned}
\Re(\text{A}_{1}) &=
\left(
\begin{array}{ccc}
a & if & 0\\
-if & a & 0\\
0 & 0 & b
\end{array}
\right),\\[2mm]
\Re(\text{E}^{+}) &=
\scalebox{0.9}{$
\left(
\begin{array}{ccc}
c & id & 0\\
-id & -c & 0\\
0 & 0 & 0
\end{array}
\right)
$},
\qquad
\Re(\text{E}^{-}) =
\scalebox{0.9}{$
\left(
\begin{array}{ccc}
id & c & 0\\
c & -id & 0\\
0 & 0 & 0
\end{array}
\right)
$}.
\end{aligned}
\label{Eqn_6m}
\end{equation}
Hence, the Raman scattering cross-section in the lab frame can be calculated for the A$_1$ modes and E modes as, $I(A_{1}(1))= |a  \sin\varphi|^2+|f  \cos\varphi|^2$, $I(E^{+})= |c \cos\varphi-d\sin\varphi|^2$, and $I(E^-)=(c+d)^2  \cos^2\varphi$. Thus, for A$_1$ modes and E mode, one expects the intensity variation for the scattered light with scattering angle to be bi-lobed with maximum either along $0^\circ$ (for the E mode) or $90^\circ$ (for the A modes) for both RCP and LCP incident light configurations.

With the above understanding of angle-resolved Raman intensity variation, we first examine the polar plot of the intensity variation of the A$_1$ and E modes of F3GT in Figure~\ref{polar_circular_F3GT}. While the intensity variation of the A$_1$ mode is bi-lobed, as expected, that of the degenerate E$_1$ modes under RCP and LCP is markedly different. It is to be recalled that for F3GT, the magnetization is predominantly oriented along the $c$ axis. Thus, the magnetization vector under rotation about the $c$-axis under in-plane RCP and LCP remains unchanged.  The in-plane Raman tensor elements remain equivalent ($\Re_{xx}=\Re_{yy}$), reflecting symmetry and polarization selection rules, leading to a helicity-symmetric Raman response for the achiral A$_1$ mode. 
In Figure~\ref{polar_circular_F3GT}(a), the blue curves are the best fit to the data points for the A$_1$ mode using the Raman tensors in Eq.~(\ref{Eqn_6m}). For the degenerate E modes, the corresponding trends are represented by the blue curves in Figure~\ref{polar_circular_F3GT}(b). It is obvious that the intensity variation of the degenerate E$^{+}$ and E$^{-}$ modes, with a four-lobed pattern in LCP and a nearly elliptical pattern under RCP, cannot be explained using the magnetic Raman tensor in Eq.~(\ref{Eqn_6m}) for F3GT. 
\begin{figure*}[t!]
\centering
	\includegraphics[width=0.95\linewidth]{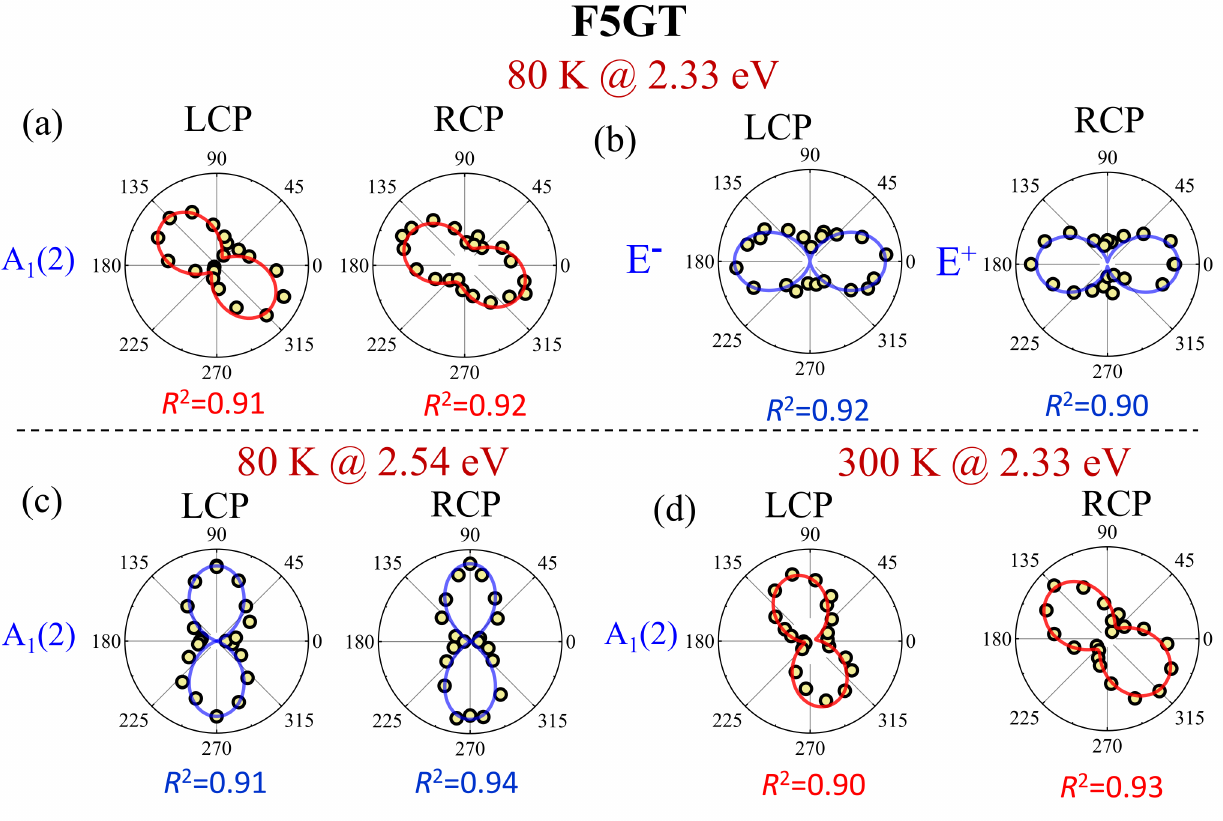}
	\caption{F5GT: Polar plots of normalized Raman intensity in LCP and RCP configurations  at 80 K (a) for A$_1$(2) mode and (b) for E(1) mode using 2.33 eV as the excitation energy. (c) Polar variation of intensity of A$_1$(2)  mode in LCP and RCP configurations at 80 K under non-resonant conditions (i.e., when spectra recorded at 2.54 eV), and (d)  the same at 300 K with 2.33 eV as the excitation energy. In all plots, data points are represented by olive solid balls, and the solid blue curves are the best fit to the data points using the tensor in Eq.~(\ref{Eqn_3m}) and red curves are the same using the tensor in Eq.~(\ref{DA1_mod_F5GT}). The corresponding values of $R^2$ are available below the plots  in blue and red fonts, respectively. The fitted parameters are available in the Sec.~S7 in SM \cite{supplementary}. }
	\label{polar_circular_F5GT}
\end{figure*}

To further establish that the observed behavior of the intensity of the E modes is indeed a resonance effect, we recorded Raman spectra of F3GT using different laser excitation wavelengths, as shown in Sec.~S6 in the SM \cite{supplementary}. The spectra are fitted using the relation in Eq.~(\ref{net intensity}). The change in the Fano parameter, which provides the coupling strength between the phonon and the electronic continuum, as a function of excitation wavelength is estimated. The drop in $1/q$ at the 2.54 eV excitation wavelength suggests that this wavelength yields a non-resonant condition. The polar plots of intensities of the E$_1$ modes, recorded using 2.54 eV as the excitation wavelength, are shown in Figure~\ref{polar_circular_F3GT}(c). The blue solid curves are the best fits to the data points using the Raman tensors in Eq.~(\ref{Eqn_6m}) for the magnetic point group F3GT, establishing that the angle-resolved intensity patterns for the E$_{1}^{+}$ and  E$_{1}^{-}$ modes in Figure~\ref{polar_circular_F3GT}(b) originate from the optical resonance effect.

As mentioned earlier, in the magnetically ordered phase, TRS breaking lifts the equivalence between the two counter-rotating components of the E mode, leading to helicity-dependent Raman tensors. Hence, the in-plane E mode carries two chiral components E$^{+}$ and E$^{-}$
of opposite angular momentum. Under RCP and LCP excitation, one component is selectively enhanced. Near resonance, the electronic states  are strongly anisotropic in the 
$ab$ plane due to the directional hopping of the itinerant Fe $d$-electrons. The electron-phonon coupling matrix element becomes tensorial and anisotropic, $g_{E^{+}}(\varphi) \neq g_{E^{-}}(\varphi)$, possibly via SOC, which couples spin, lattice and orbital degrees of freedom. Under resonance, this anisotropy is amplified, as the Raman process is dominated by specific electronic transitions rather than structural symmetry-averaged ones; and the polar plots of the E$_1$ mode in F3GT exhibit a pronounced helicity dependence, with distinct angular intensity patterns under RCP and LCP illumination, as observed in Figure~\ref{polar_circular_F3GT}(b).  
As the easy axis of magnetization is along the $c$-axis for F3GT, it is likely that the observed effect arises due to the convolution of both the weak in-plane effect and the indirect effect (e.g., SOC) of the magnetization vector along the $c$-axis.
The above argument is also supported by the fact that at 300 K, when the system is in the PM phase, the polar plots of the intensity of the E$^+$ and E$^-$ a modes are very similar [Figure~\ref{polar_circular_F3GT}(d)], following the Raman tensors described in Eq.~(\ref{Eqn_6m}).
\begin{figure*}
\centering
	\includegraphics[width=0.85\linewidth]{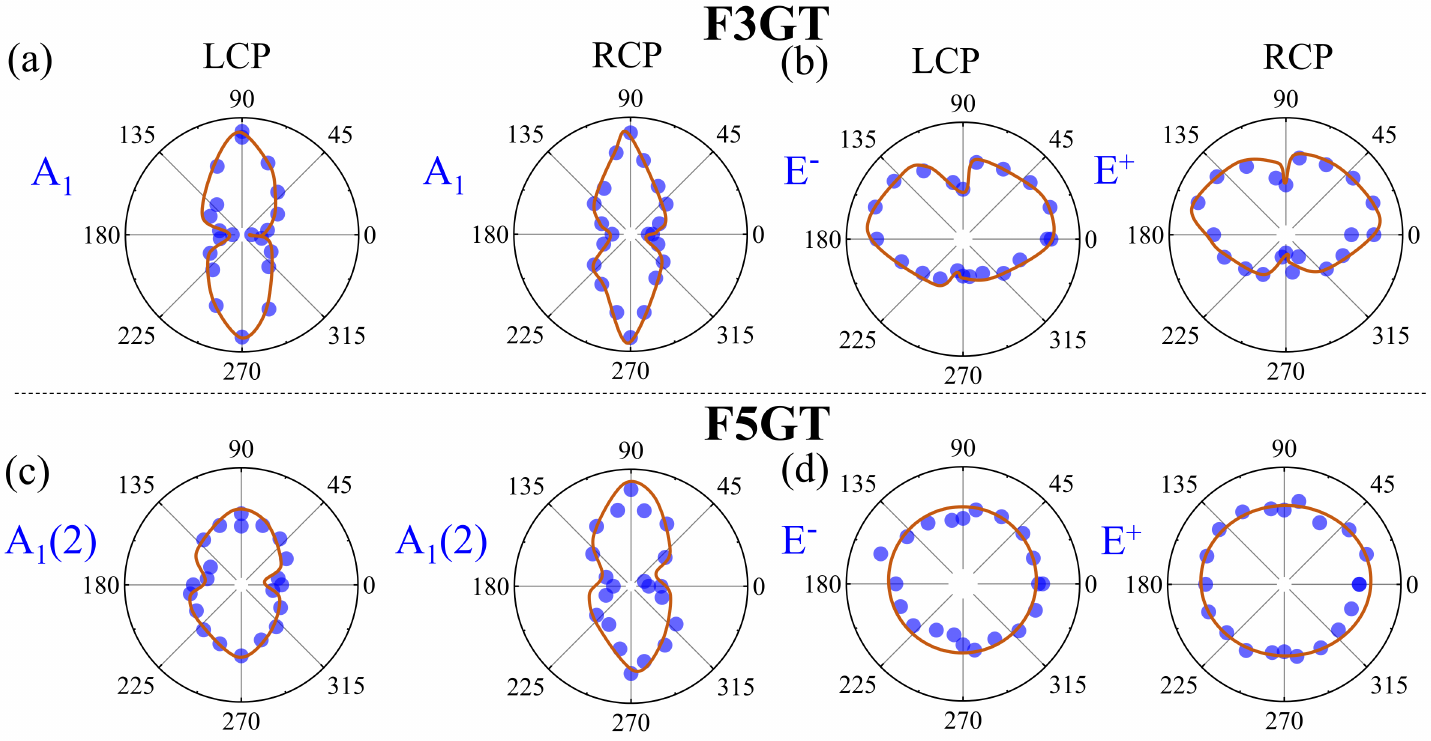}
	\caption{Angular variation of $1/q$ with incident light with RCP and LCP for (a) the A$_1$ mode and (b) for the E mode of F3GT. (c)--(d) The same for the A$_1$(2) mode and E mode, respectively, for F5GT. The solid red curves are the guide to the eyes.}
	\label{dw_t}
\end{figure*}

Figure~\ref{polar_circular_F5GT}(a)--(b) present the polar plots of the intensity of the  A$_1$(2), and E(1) modes of F5GT for varying scattering directions for the incident RCP and LCP light at 80 K. 
F5GT with $R3m$ space group belongs to the magnetic point group 3$m$; and the corresponding Raman tensor can be written as~\cite{cracknell1969};
\begin{eqnarray}
\label{eq.raman_tensor}
\nonumber \Re(\text{A}_{1}) &=& \left( \begin{array}{ccc}
    a & if & 0 \\
     -if & a & 0 \\
     0 & 0 & b \\
    \end{array}\right), \\
\nonumber \Re(\text{E}^\text{+}) &=& \left( \begin{array}{ccc}
    d & id & ic \\
     id & -d & c \\
     ig & g & 0 \\
    \end{array}\right), 
\nonumber \Re(\text{E}^-) = \left( \begin{array}{ccc}
    e & -ie & -ij \\
     -ie & -e & j \\
     -ih & h & 0 \\
     \end{array}\right). \\
     \label{Eqn_3m}
\end{eqnarray}
Using the above tensors, the intensity variation of the A$_1$ and E modes are $I(A_{1})= |(a  \sin\varphi|^2$, and $I(E^+)= 0$, $I(E^{-})= |e  \cos\varphi|^2$. 
Interestingly, for F5GT, the angle-resolved intensity variation of the E(1) modes under RCP and LCP exhibits similar behavior, and could be fitted well using the Raman tensor for the 3$m$ point group available in Eq.~(\ref{Eqn_3m}).
This suggests that while the easy axis of the magnetization modifies the phonon energies through TRS breaking and chiral spin--orbit interaction (as discussed for Figure~\ref{RL_LR}), the in-plane electron--phonon coupling remains effectively directionally isotropic; possibly due to multi-band electronic structure, interlayer hybridization, and domain averaging in F5GT~\cite{fujita2022,may2019a,may2019b}, which suppresses directional anisotropy in the Raman matrix elements, in contrast to F3GT, where anisotropic coupling leads to four-lobed or elliptical Raman responses.

Here, it is to be noted that the two observables, the frequency splitting ($\Delta f$) and angle-resolved intensity variation of the E modes under RCP and LCP light, probe two different physics channels. While the former arises due to the real part of the phonon self-energy, the latter depends on the Raman tensor amplitude and phase, which in turn depends on various factors, e.g., the symmetry of optical matrix elements, anisotropy of electron--phonon coupling, phase structure of the Raman tensor, and chirality-selective electronic transitions. While the magnetic order strongly renormalizes phonon energies and results in larger splitting, the increased Fe-layer complexity, multiple inequivalent Fe sites, and enhanced interlayer hybridization in F5GT \cite{Yamagami2022} lead to an effective averaging of chirality-dependent coupling, resulting in nearly identical RCP and LCP, $g_{E^{+}}(\varphi) \approx g_{E^{-}}(\varphi)$ polar responses. This suppresses directional anisotropy in the Raman matrix elements, unlike F3GT, which exhibits anisotropic coupling due to 
enhanced interlayer hybridization .

However, for F5GT,  A$_1$(2) mode shows a tilt at 80 K in the intensity plot for both RCP and LCP configurations. For the achiral A$_1$, the helicity-independent nature of this tilt indicates that it arises from static structural anisotropy rather than magneto-chiral effects. Considering the damped oscillator model for an electrical dipole, the susceptibility is a complex number near resonance~\cite{pimenta2021}. Hence, the Raman tensors can be expressed in a complex polar plane, with imaginary matrix elements and a relative phase between them. Manifestation of such a complex relative phase in Raman intensity, thus, carries the signature of electron-phonon coupling in the system under resonance. Also refer to our previous article \cite{mekap2026} on linearly-polarized Raman measurements on F5GT for further details.  Hence, using complex tensor elements, the intensity plot could be best fitted by considering the Raman tensor
\begin{eqnarray}
\nonumber \Re A_1({mod}) = \left( \begin{array}{ccc}
    a \exp \left( {i\alpha_{a}} \right) & if & 0 \\
     -if & a \exp \left( {i\alpha_{a^{\prime}}} \right) & 0 \\
     0 & 0 & b \exp \left( {i\alpha_{b}} \right) \\
   \end{array}\right) . \\
   \label{DA1_mod_F5GT}
\end{eqnarray}
Hence, the calculated Raman intensity becomes:
\begin{eqnarray}
\nonumber I(\Re A_{1}(mod)) &=& | a (\cos \varphi+\sin{\varphi}) - 2f\sin\varphi\cos\alpha_{a} \\
\nonumber && - a\cos\varphi \cos\alpha_{a^{\prime}a} +a\sin\varphi \sin\alpha_{a^{\prime}a} |^2 \\
\nonumber && + | 2f\sin\varphi\sin\alpha_{a}-a\cos\varphi \sin\alpha_{a^{\prime}a} \\
&& + a\sin\varphi\sin\alpha_{a^{\prime}a} |^2
\label{Int_DA1mod_F5GT}
\end{eqnarray}
Here, $\alpha_{a^{\prime}a}=\alpha_{a^{\prime}}-\alpha_a$, is the difference between the phase angles in the tensor elements. 
The phase factors indicate the relative time responses of the electronic polarization induced by lattice vibrations in different directions. The fitted parameters are available in  Sec.~S7 of SM~\cite{supplementary}. Here we would like to mention that in the angle-resolved linear-polarization measurements in Ref. \cite{mekap2026}, the incident polarization was kept fixed while the analyzer is rotated by $\theta$. For this geometry, the measured Raman amplitude depends on the tensor elements associated with the fixed incident polarization, while the phase of the orthogonal diagonal element is not independently constrained, unlike the helicity-dependent Raman measurements. We therefore used $\alpha_{a'}=\alpha_a$ in describing the linear-polarization data, without interpreting this equality as an independent experimental determination of the two phases.
 It is to be noted that A$_1$ modes couple to the scalar electronic density and are hence sensitive to phase interference, giving rise to a static phase offset. A similar tilt in the polar plot of the Raman intensity of the A$_1$ mode has been reported earlier~\cite{mekap2026} using linear-polarization Raman scattering and has been attributed to optical-resonance-induced electron-phonon coupling in the system.
It is noteworthy that the effect vanishes under off-resonance conditions. As in the case of F3GT, the Raman spectra of F5GT are recorded at different excitation wavelengths, and the off-resonance excitation wavelength of 488 nm is chosen from the $1/q$ plot (see Sec. S6 in the SM \cite{supplementary}). The intensity variation of the A$_1$ mode could be explained without introducing the phase factor in the tensor elements in Eq.~(\ref{DA1_mod_F5GT}) in Figure~\ref{polar_circular_F5GT}(c). Nonetheless, it is interesting to note that the tilt in the polar plot of the intensity also persists at 300 K, as shown in Figure~\ref{polar_circular_F5GT}(d), when the system is in the PM phase. Above $T_C$, short-range spin correlation exists in the absence of long-range magnetic order. The spin-orbit coupling mediated short-range spin correlation results in an effective modification of Raman susceptibility, which, in turn, causes symmetry mixing \cite{kong2024,liu2023}.  The tilt in A$_1$ mode has been explained by the dynamical mixing of A$_1$ and E symmetry channels mediated by spin-orbit coupled spin fluctuations in PM phase \cite{mekap2026}.

The above discussion reveals strong electron-phonon coupling in both systems in the magnetically ordered phase,  mediated either by magnetic anisotropy or static structural identity.  The same could be further revealed by analyzing the angular variation of the Fano parameter $1/q$ at 80 K.
The Fano parameter $q$ in Eq.~(\ref{net intensity}) reflects the nature of the phase and the interference of the discrete phonon with the electronic continuum of excitations. The coupling strength is given by $1/q$. Figure~\ref{dw_t}(a)--(d) plot the angle-resolved values of  $1/q$ at 80 K in both RCP and LCP configurations. The anisotropic nature of in-plane electron-phonon-spin coupling strength has been revealed for F3GT in Figure~\ref{dw_t}(a) and (b), as discussed earlier. While the out-of-plane electron-phonon coupling via the A$_1$ mode is affected by easy-axis magnetization, the in-plane E$_1$ mode sees the indirect consequence of the magnetization vector. We note the asymmetry of the polar plot of $1/q$ of the E$_1$ mode about the horizontal axis. As for the A$_1$ mode of the same plot being symmetric and obtained from the same recorded spectra at an angle, we argue that the observed asymmetry for the E$_1$ mode is not an experimental artifact due to the misalignment of the polarizer or imperfect quarter-waveplate calibration. The asymmetric plot in Figure~\ref{dw_t}(b) possibly implies
magnetization-induced antisymmetric Raman tensor
coupling between chiral phonon and SOC-driven orbital response; the same as suggested by the polar plot of the Raman intensity of the E$_{1}^{+}$ and E$_{1}^{-}$ modes in Figure~\ref{polar_circular_F3GT}(b). In F5GT, the isotropic nature of the angular variation of the $1/q$ plot of the E(1) mode confirms the complexity in electron-phonon coupling due to the increased Fe-layer, multiple inequivalent Fe sites, and enhanced interlayer hybridization in F5GT.

\begin{figure*}
\centering
	\includegraphics[width=0.95\linewidth]{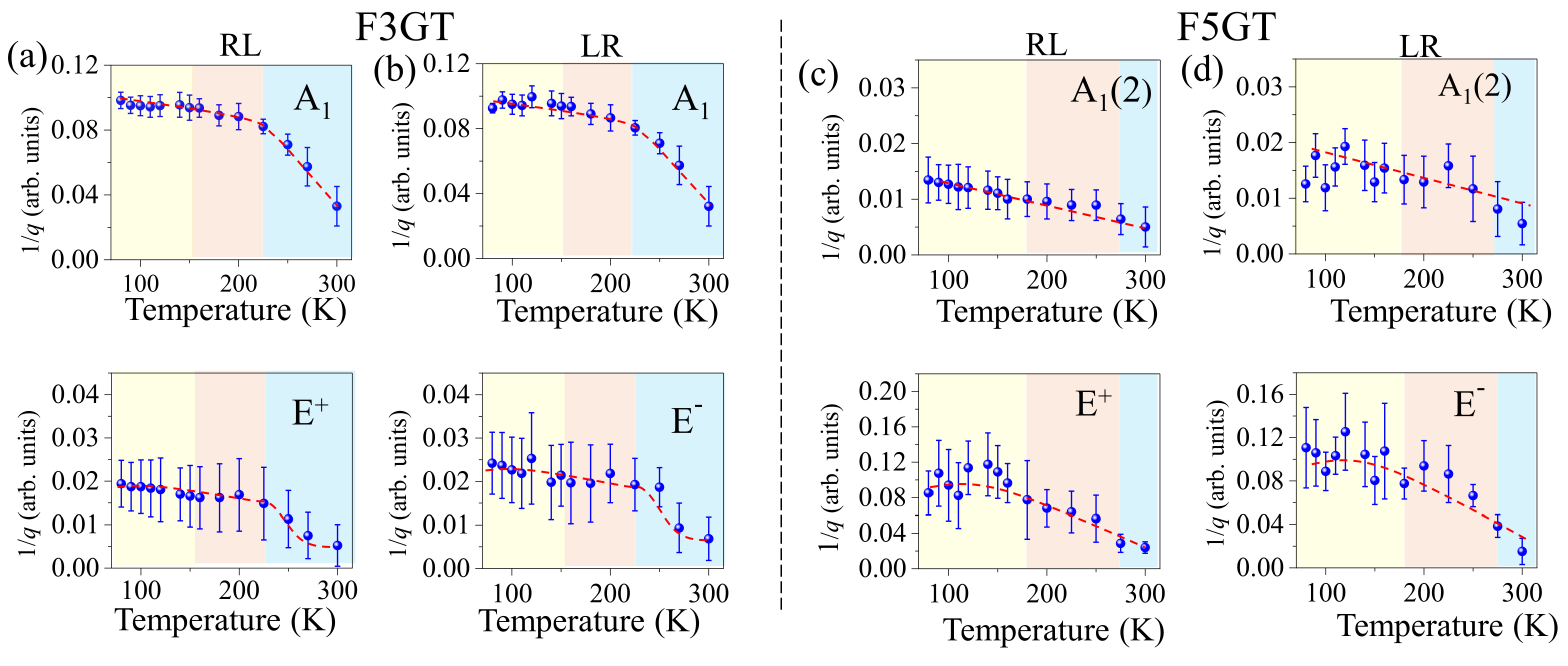}
	\caption{Evolution of $1/q$ with temperature for F3GT in (a) RL and (b) LR configurations for the A$_1$ and E modes. (c)--(d) The same for F5GT, respectively, for the A$_1$(2) and E modes. The dashed curves in all the panels are guides to the eyes. The error bars to the data point are estimated from the standard deviation of the fitted value of $q$. The color shades mark three regimes of distinct spin-correlations as in Fig.~\ref{RL_LR}.}
	\label{fano_temperature}
\end{figure*}

Figure~\ref{fano_temperature}(a)--(d) exhibit the variation of 1/$q$ with temperature for both F3GT and F5GT in LR and RL configurations.  In all panels of the figure, the green shaded area marks the magnetically ordered phase. Except for the E$_1$ mode in F3GT, for all other modes in LR and RL configurations, 
the electron-phonon coupling strength, i.e., the $1/q$ parameter, decreases monotonically for both compounds with the rise in temperature. However, the same for E$_1$ mode of F3GT exhibits a nonmonotonic trend below and above $T_C$. Above $T_C$, the electron-phonon coupling parameter, $1/q$, drops much more sharply. This behavior once again suggests that the electronic continuum responsible for the Fano interference is strongly mediated by magnetic ordering in F3GT, with the effective electron--phonon coupling enhanced by spin polarization in the FM phase. Such coupling collapses when time-reversal symmetry is restored.

\subsection{Helicity dependent light-phonon-electron-spin response}

In helicity-resolved Raman measurements, the Raman asymmetry parameters provide a sensitive measure of the anisotropy and dynamical character of phonon-induced electronic and spin polarizability, capturing the influence of electron-phonon-spin coupling, itinerant electrons, and magnetic ordering beyond static symmetry considerations~\cite{lyu2020,yin2021}.  

We define the co-circular Raman asymmetry parameter as
\begin{equation}
\rho = \frac{I(RR) - I(LL)}{I(RR) + I(LL)},
\end{equation}
which indicates the optical preference of the system for circularly polarized light, and hence, is sensitive to the magnetic optical activity of the chiral modes.\\
We also define the cross-circular normalized helicity asymmetry parameter as
\begin{equation}
\rho_{1} = \frac{I(RL) - I(LR)}{I(RL) + I(LR)},
\end{equation}
which serves as an order-parameter-like quantity associated with TRS breaking and complex electron-spin-phonon coupling in FGT. 
$\rho_1$ provides the ability of the system for helicity-converting light-matter interaction.

\begin{figure*}
\centering
\includegraphics[width=0.9\linewidth]{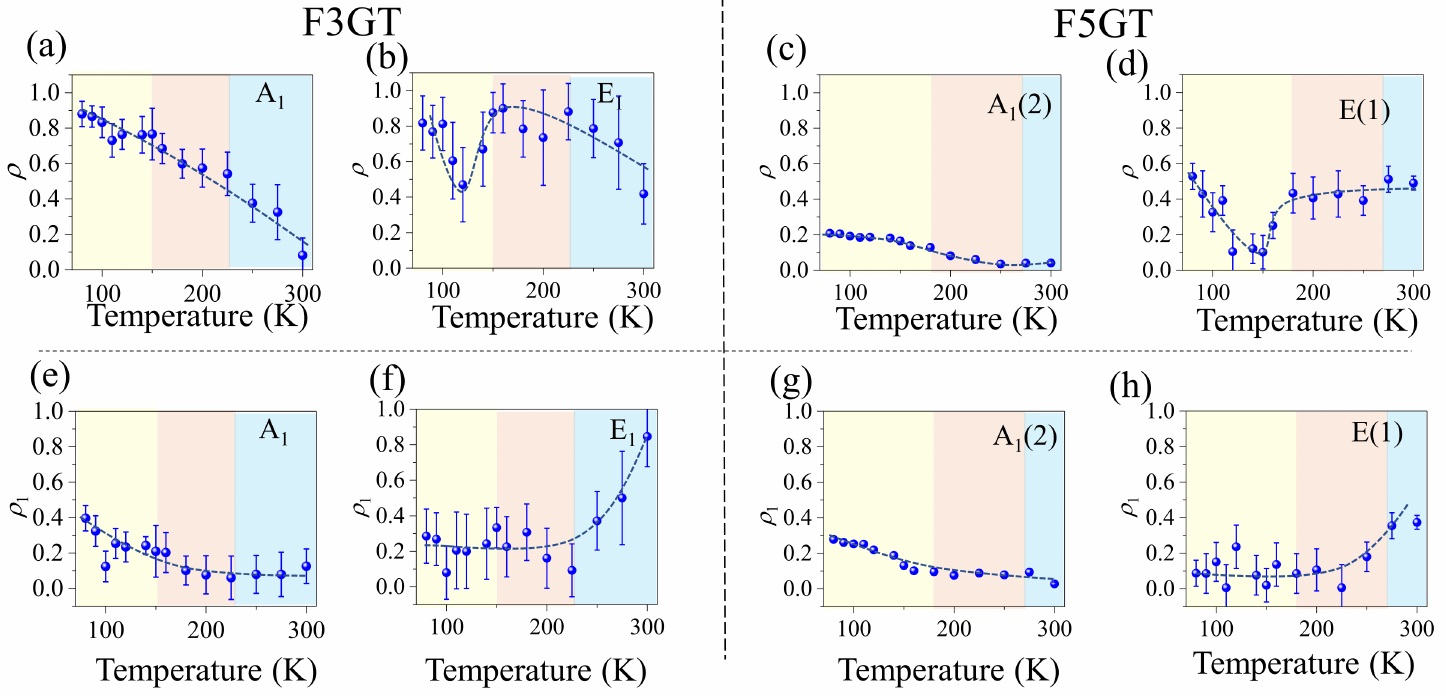}
	\caption{Evolution of  co-circular Raman asymmetry parameter ($\rho$) with temperature for (a)--(b) F3GT,  and (c)--(d) F5GT. Temperature variation of cross-circular Raman asymmetry parameter, $\rho_1$ in  (e)--(f) F3GT, and  (g)--(h) F5GT. The dashed curves in all the panels are guides to the eyes. The error bars to the data point are estimated from the standard deviation of the fitted value of Raman intensities of the corresponding modes in different polarization configurations. The color shades mark three regimes of distinct spin-correlations as in Fig.~\ref{RL_LR}.}
	\label{depolarization}
\end{figure*}

Figure~\ref{depolarization}(a)-(d) plot the $\rho$ of all modes under study over the temperature range of 80 to 300 K for both F3GT and F5GT.
We observe a markedly different temperature evolution of the $\rho$ for A$_1$ and E modes, highlighting the mode-selective nature of electron-phonon-spin coupling in these systems to circularly polarized light. For both F3GT and F5GT the out-of-plane A$_1$
 modes track the gradual buildup of magnetic order along the 
$c$-axis with temperature.
The dip in the parameter $\rho$ around 120 K for the E modes suggests a reduction in the difference between the RR and LL Raman intensities. The anomalous behavior of the E phonon in the low-temperature regime possibly arises as they  are sensitive to in-plane symmetry breaking, and  reflects a reorganization of the underlying magnetic correlations rather than a simple reduction of the magnetic order parameter. The absence of a similar feature for the A$_1$ mode further suggests that the effect originates from the evolution of anisotropic spin correlations, associated with the competition between itinerant and localized magnetism \cite{Saini2026} or changes in magnetic domain configurations at low temperatures \cite{chen2023}.

Figure~\ref{depolarization}(e)--(h) plot the evolution of $\rho_1$ with the temperature of the systems. For both F3GT and F5GT, we observe a gradual decrease in $\rho_1$ with increasing temperature for the A$_1$ modes. It goes to zero above $T_C$. For the achiral A$_1$ mode, the helicity asymmetry mainly arises due to magnetization-induced antisymmetric Raman tensor terms. Consequently, the temperature variation of $\rho_1$ for the A$_1$  modes describes an order-parameter-like behavior governed predominantly by the static magnetic order, as observed for $\rho$.
However, for the E modes, the trend is the opposite. Such a strong selective nature suggests that light-matter interaction is not solely governed by magnetization but by the symmetry-dependent coupling between phonons, spin polarization, and electronic states, as reflected in the temperature variation of $\rho_1$ of the E modes.  The intensity of chiral E modes depends on various factors—PAM of the phonon, 
spin-orbit coupling, and
magnetic fluctuations, magnetic itinerancy discussed earlier.
While the static magnetization decreases, magnetic fluctuations may increase strongly; and the E mode can couple with
transverse spin fluctuations and 
fluctuating local chirality
dynamic via spin-orbit coupling.  Hence, it reflects strong spin-orbit-entangled anisotropic electron-phonon coupling in FGT.

 These results establish the helicity-resolved Raman asymmetry parameters as a powerful window to probe the interplay between lattice dynamics, itinerant electrons, and magnetic correlations, particularly in the presence of anisotropic exchange interactions.

\section{Summary}
In summary, comparative helicity‑resolved Raman spectroscopy reveals fundamentally distinct anisotropic electron-phonon-spin coupling mechanisms in F3GT and F5GT. Despite both compounds exhibiting chiral E mode splitting under circularly polarized light, a signature of time‑reversal symmetry breaking, the temperature dependence of this splitting ($\Delta f$) does not track the magnetization in either system. This mismatch demonstrates that spin‑orbit‑coupled electronic interactions play an essential, previously underappreciated role in the magneto‑chiral response, beyond simple magnetic-order-driven TRS breaking.

The two compounds diverge in their in-plane anisotropic coupling: F3GT exhibits resonance‑induced, magnetization‑mediated anisotropy, resulting in four‑lobed and elliptical polar patterns for E$^+$ and E$^-$ modes, respectively. In contrast, F5GT shows static, helicity-symmetric polar responses due to its multiple inequivalent Fe sites and enhanced interlayer hybridization, which average out directional anisotropy. The Fano asymmetry parameter $1/q$ quantitatively captures these differences: strongly anisotropic in F3GT but isotropic in F5GT. The cross-circular asymmetry parameter $\rho_1$ serves as an effective order parameter for TRS breaking, vanishing above $T_C$ for both A$_1$ and E modes.

These findings establish that tuning Fe stoichiometry and layer stacking in Fe$_n$GeTe$_2$ compounds offers a powerful handle for engineering electron‑phonon‑spin interactions – from resonance‑mediated dynamical anisotropy to static, symmetry‑averaged coupling. Functionally, the strong helicity‑dependent anisotropy in F3GT suggests its potential for low‑power spin‑orbit torque switching assisted by chiral phonons, where the four‑lobed Raman response indicates deterministic all‑optical control of the magnetization direction. Conversely, the isotropic electron‑phonon coupling and large chiral splitting in F5GT make it an ideal platform for chiral magnonic waveguides with non‑reciprocal spin‑wave propagation. An open question remains whether these contrasting behaviors persist in the ultrathin limit, where reduced dimensionality may enhance or suppress the observed anisotropies.

\begin{acknowledgments}
Some figures in this work were rendered using {\sc Vesta}~\cite{momma2011} software. SG thanks Anusandhan National Research Foundation, Government of India (ANRF/PAIR/2025/000029/PAIR-A) for financial assistance for this project. AR acknowledges the aids from Overleaf Writeful Premier for language editing. 
\end{acknowledgments}

\section*{Data Availability}
Data supporting the findings of this study are available from the corresponding author on a reasonable request.

\clearpage
\newpage

\onecolumngrid
\begin{center}
  \textbf{\Large Supplemental Material}\\[.5cm]
  \textbf{\large Contrasting anisotropic electron-phonon-spin coupling in Fe$_{3}$GeTe$_{2}$ and Fe$_{5}$GeTe$_{2}$: A helicity-resolved Raman study}\\[.3cm]
  Smrutiranjan~Mekap$^{1}$,  Jyoti~Saini$^{2}$, 
  Andrzej~Ptok$^{3}$,
 Pawan Kumar Srivastava $^{3}$, Changgu Lee $^{3,4}$, Subhasis~Ghosh$^{2}$, and Anushree~Roy$^{1}$ \\[.2cm]
  {\itshape
${}^{1}$Department of Physics, Indian Institute of Technology Kharagpur, Kharagpur 721302, India \\[.1cm]
${}^{2}$School of Physical Sciences, Jawaharlal Nehru University, New Delhi-110067, India\\[.1cm]
${}^{3}$Institute of Nuclear Physics, Polish Academy of Sciences, ul. W. E. Radzikowskiego 152, 31-342 Krak\'{o}w, Poland\\[.1cm]
${}^{3}$ School of Mechanical Engineering, Sungkyunkwan University, Suwon 16419, Republic of Korea.\\[.1cm]
${}^{4}$ SKKU Advanced Institute of Nanotechnology (SAINT), Sungkyunkwan University, Suwon 16419, Republic of Korea.}

  (Dated: \today)
\\[0.3cm]
\end{center}

\setcounter{equation}{0}
\renewcommand{\theequation}{S\arabic{equation}}
\setcounter{figure}{0}
\renewcommand{\thefigure}{S\arabic{figure}}
\setcounter{section}{0}
\renewcommand{\thesection}{S\arabic{section}}
\setcounter{table}{0}
\renewcommand{\thetable}{S\arabic{table}}
\setcounter{page}{1}

\begin{figure*}[htbp]
\centering
	\includegraphics[width=\linewidth]{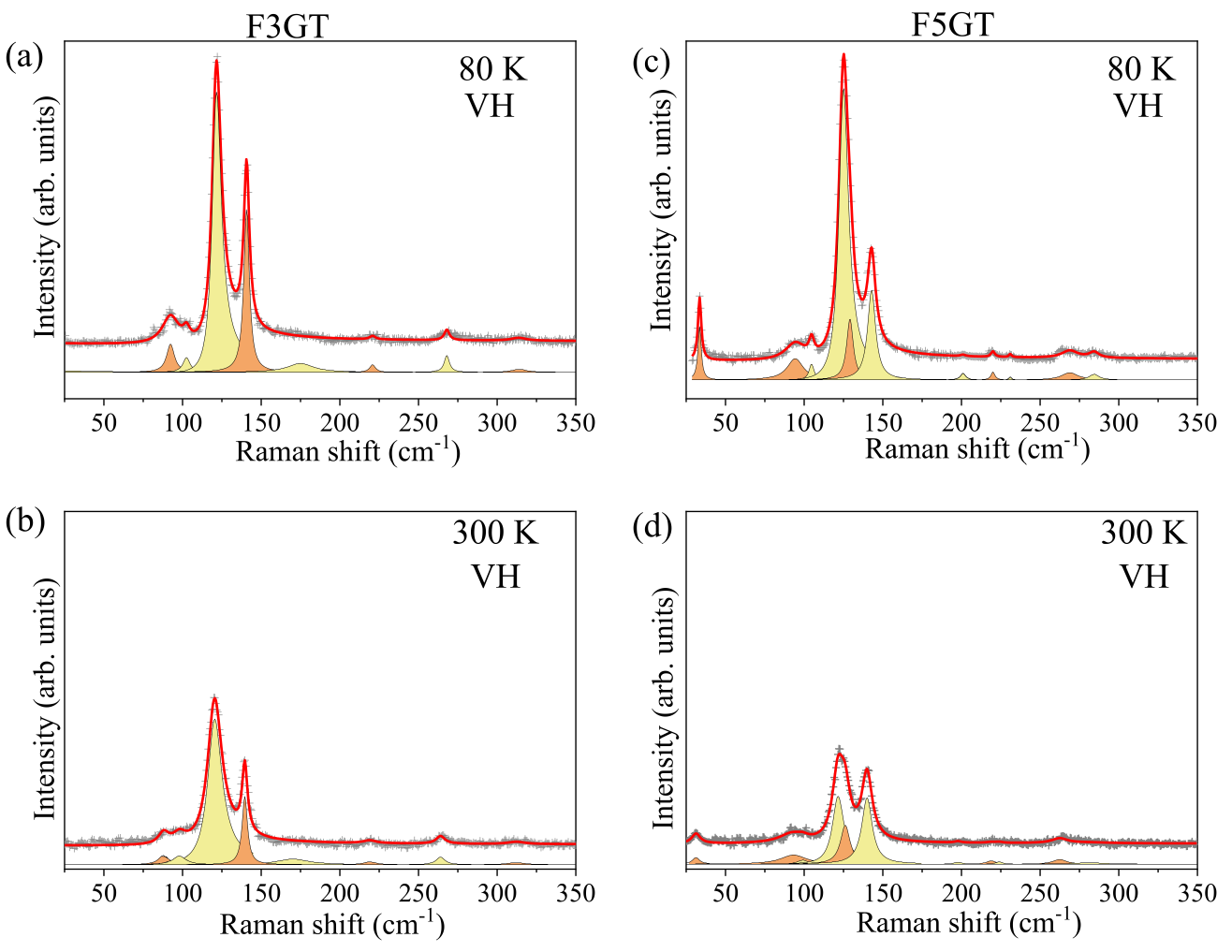}
	\caption{(Color online) Background-subtracted Raman spectrum measured at (a) 80 K, and (b) 300 K for F3GT in VH  in polarization configuration, and the same for F5GT in (c) and (d).}
	\label{fig.linear.vh}
\end{figure*}

In this Supplemental Material, we present additional results:
\begin{itemize}
\item Sec.~\ref{sec.raman spectra} Raman spectra of F3GT and F5GT
\begin{itemize}
    \item Fig.~\ref{fig.linear.vh} Raman spectra of F3GT and F5GT at 80 K and 300 K.

\end{itemize}
\item Sec.~\ref{sec.F3GT_F5GT_TeO2_Te} -- Intrinsic origin of the Raman spectra of F3GT and F5GT.
\begin{itemize}
    \item 
  A. Fig.~\ref{fig:Te_TeO2_specta}  Comparison of Raman spectra of TeO$_2$, Te, F3GT and F5GT. 
\end{itemize}
\begin{itemize}
    \item B. Fig.~\ref{fig:F5GT_DFT}  DFT-derived atomic displacement vector.
\end{itemize}
\begin{itemize}
    \item C. Fig.~\ref{fig:cleaved_FGT}  Raman spectra of cleaved and uncleaved F3GT and F5GT.
\end{itemize}

\begin{itemize}
    \item D. Fig.~\ref{fig:wavelength_Te}  Wavelength-dependent Raman intensity of TeO$_2$, Te, F3GT, and F5GT.
\end{itemize}
\begin{itemize}
    \item E.  Systematic differences in Raman spectral behavior of F3GT and F5GT.
\end{itemize}
\begin{itemize}
    \item F. Existence of chiral phonon modes.
\end{itemize}

\begin{itemize}
    \item G. Fig.~\ref{fig:power_wavelength}  Similar spectral profile for different laser powers and different laser wavelengths.
\end{itemize}

\item Sec.~\ref{sec.background} -- Raman background continuum
\begin{itemize}
    \item A. Fig.~\ref{fig:bg_spectra} -- Raman background continuum of F3GT and F5GT.
    \item B. Fig.~\ref{fig:bg_wavelength} --Wavelenth-depenent Raman background continuum of F3GT and F5GT.
    \item 
    C. Fig.~\ref{fig:bg_TeO2}--Comparison of background continuum of FGT with TeO$_2$
    
\item D. Fig.~\ref{Fano_fit} --Statistically consistent description for Fano function.
 \end{itemize}
\item Sec.~\ref{sec.symmetry} -- Symmetry assignments of Raman modes
\begin{itemize}
\item Fig.~\ref{fig:linear_symmetry} -- Symmetry assignments for Raman modes.
\end{itemize}
\item Sec.~\ref{sec.tempr_depnd} -- Temperature-dependent Raman studies under RL and LR configurations

\begin{itemize}
\item Fig.~\ref{fig:splitting_fgt} -- Temperature-dependent Raman studies under RL and LR polarization configurations.
\end{itemize}
\item Sec.~\ref{sec.wavelength_depnd} --Wavelength- dependent Raman measurements
\item Sec.~\ref{sec.fitted_parameter} -- Fitted parameters for F5GT in Fig. 5 using Eqn. 5 of the main text
\begin{itemize}
    \item Table.~\ref{fitted_parameter}  Fitted parameters for F5GT in Fig. 5 using Eqn. 5 of the main text.
\end{itemize}
\end{itemize}
\newpage
\section{Raman spectra of F3GT and F5GT at 80 K and 300 K}
\label{sec.raman spectra}

Spectra are recorded using cross-polarized VH ($z(xy)\bar{z}$, $z$ along $c$-axis, $x$ along $a$-axis, and $y$ along orthogonal $b$ axis of the crystal) configuration. In both panels, the background-subtracted spectra are shown by + symbols. Spectra are analyzed following Eq.~1 of the main text. The modes with A and E symmetries are shown by yellow and orange shaded areas. Red curves represent the net fitted spectra.

\newpage
\section{Intrinsic origin of the Raman spectra of F3GT and F5GT}
\label{sec.F3GT_F5GT_TeO2_Te}
\subsection{Comparison of Raman spectra of TeO$_2$, Te, F3GT and F5GT} \label{Raman_spectra}
In the literature, there is concern about possible TeO$_2$ or elemental Te surface degradation products in F3GT and F5GT, as some of the peaks are apparently similar in all four compounds.  Below, we provide detailed supporting experimental evidence showing the intrinsic origin of the spectral profile of F3GT and F5GT in this article.

\begin{figure}[h]
    \centering
    \includegraphics[width=0.8\linewidth]{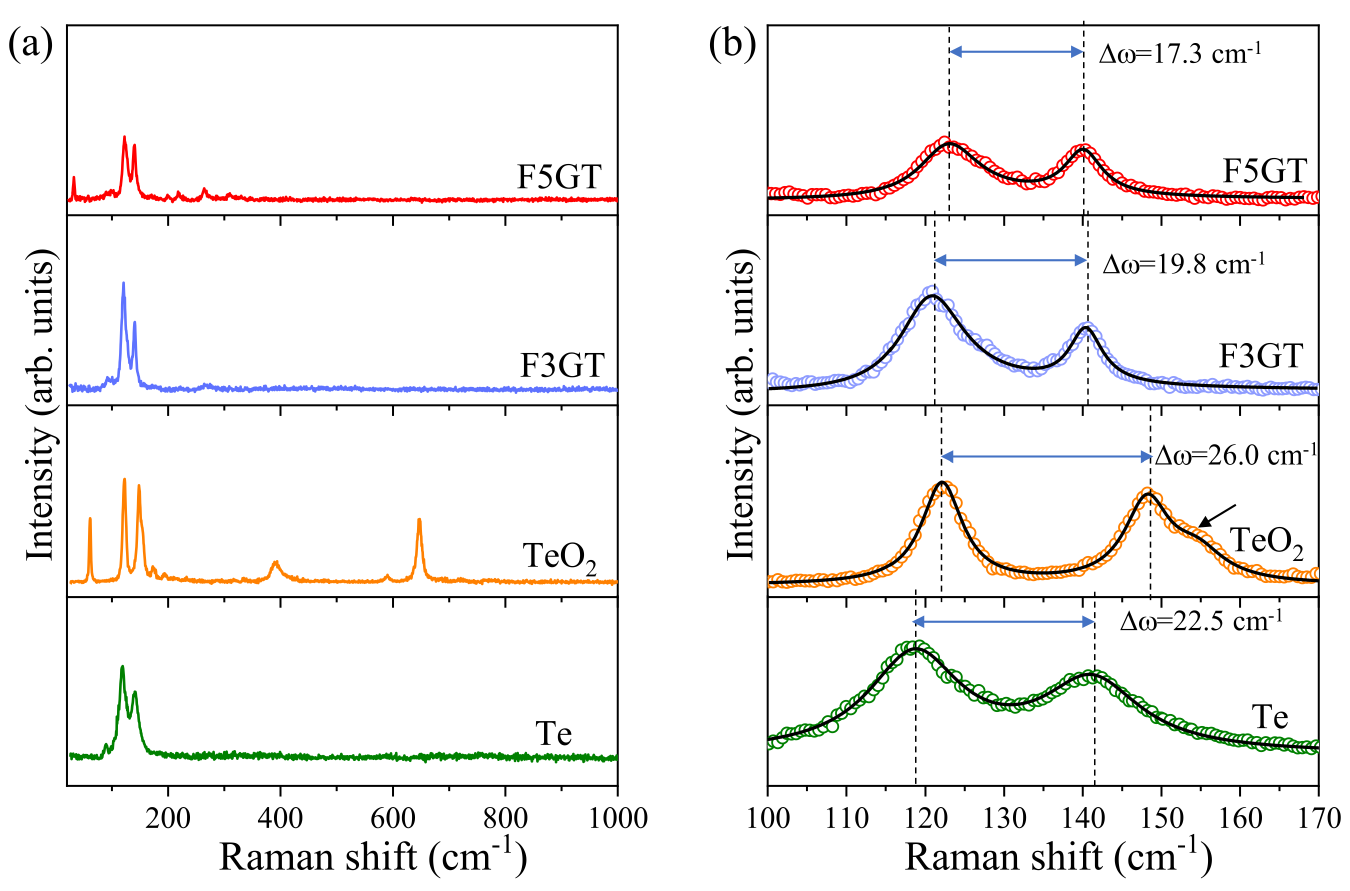}
    \caption{ (a) Comparison of the room temperature unpolarized fitted- background-subtracted Raman spectra of F3GT and F5GT, and linear background subtracted spectra of Te, and TeO$_2$ over the spectral range between 20 and 1000 cm$^{-1}$. (b) Plots the magnified Raman spectra of the same over the spectral range between 100 and 170 cm$^{-1}$.} 
    \label{fig:Te_TeO2_specta}
\end{figure}
 \noindent Refer to Fig. \ref{fig:Te_TeO2_specta}. Below we list the characteristic Raman peaks of $\alpha$-TeO$_2$ (paratellurite), elemental Te, F3GT, and F5GT. \\
From Fig. \ref{fig:Te_TeO2_specta}, Raman shift:\\
For TeO$_2$: 61.3 cm$^{-1}$, 122.2 cm$^{-1}$, 148.2 cm$^{-1}$, 154.3 cm$^{-1}$, 172.5 cm$^{-1}$, 193.4 cm$^{-1}$, 209.6 cm$^{-1}$,316.0 cm$^{-1}$, 334.0 cm$^{-1}$, 390.95 cm$^{-1}$, 590.2 cm$^{-1}$, 646.4 cm$^{-1}$.\\
For Te: 89.4 cm$^{-1}$, 118.7 cm$^{-1}$, 141.1 cm$^{-1}$.\\
For F3GT: 87.5 cm$^{-1}$, 98.5 cm$^{-1}$, 120.6  cm$^{-1}$ (A$_1$ mode), 139.6 cm$^{-1}$ (E mode), 169.66 cm$^{-1}$, 218.9 cm$^{-1}$, 265.1 cm$^{-1}$, 311.9 cm$^{-1}$ .\\
For F5GT: 31.3 cm$^{-1}$ (E mode), 92.2 cm$^{-1}$, 99.1 cm$^{-1}$, 122.5 cm$^{-1}$ (A$_1$ mode), 126.0 cm$^{-1}$ (E mode), 140.7 cm$^{-1}$ (A$_1$ mode), 197.8 cm$^{-1}$, 218.9 cm$^{-1}$, 224.1 cm$^{-1}$, 262.5 cm$^{-1}$, 281.4 cm$^{-1}$. \\

It is to be noted that the Raman spectrum for TeO$_2$ shows low-frequency modes (61.3 cm$^{-1}$, 122.2 cm$^{-1}$, 148.2), and also a high-frequency mode at 390.95 cm$^{-1}$, 646.4 cm$^{-1}$, which are consistent with reports [Smirnov et al. J. Phys. Condens. Matter 30,  475403 (2018)]. The elemental Te exhibits characteristic modes at  89.4 cm$^{-1}$, 118.7 and 141.1 cm$^{-1}$, consistent with the reported Raman modes of Te [Manjón et al., J. Mater. Chem. C 9, 6277–6289 (2021)]. A weak shoulder at $\sim$154.3 cm$^{-1}$ for TeO$_2$ is consistent with the literature.  
It is also to be noted that the presence of strong  $\sim$646 cm$^{-1}$  and $\sim$ 122 cm$^{-1}$  is a hallmark of Raman peaks of TeO$_2$  [Smirnov et al. J. Phys. Condens. Matter 30,  475403 (2018), which are missing in the Raman spectra of F3GT or F5GT. Withing the narrow spectral between 90 and 150 cm$^{-1}$, the Raman spectrum of Te is very similar to that of F3GT/F5GT. However, there are additional Raman modes in F3GT/F5GT, which are missing in the Raman spectrum of Te. In F5GT, we find an additional peak at 34 cm$^{-1}$, which is neither present in TeO$_2$, nor in Te or F3GT. \\
 \\
In Fig. \ref{fig:Te_TeO2_specta}(b), we compare a magnified view over the spectral range between 100 and 170 cm$^{-1}$. The separation between two strong peaks ($\Delta\omega$) over this spectral range is clearly different in F3GT/F5GT and TeO$_2$/Te. \\
$\Delta\omega$ for TeO$_2$: 26.0 cm$^{-1}$\\
$\Delta\omega$ for Te: 22.5 cm$^{-1}$\\
$\Delta\omega$ for F3GT: 19.8 cm$^{-1}$\\
$\Delta\omega$ for F5GT: 17.3 cm$^{-1}$\\
Here we would like to note that the resolution of the triple monochromator spectrometer is  $\sim$0.20 cm$^{-1}$, for our experimental configuration, which includes slit width, dispersion of the grating for 532 nm as the excitation wavelength, and the area of the CCD detector in pixels.\\
Moreover, the right shoulder peak  (marked by an arrow in Fig. \ref{fig:Te_TeO2_specta}(b))  at $\sim$ 154.3 cm$^{-1}$ for TeO$_2$ is missing in FGT samples. In the sharper spectra of single crystals of the FGT sample, this peak would have been more prominent if the origin of the spectra were TeO$_2$. 
All spectra reported in Fig. \ref{fig:Te_TeO2_specta} are recorded under identical experimental configuration and integration time. Hence, the intensity scale is kept deliberately the same. Please refer to Fig. \ref{fig:F5GT_DFT} for a larger view of F3GT and F5GT spectra.

\subsection{DFT-derived atomic displacement vectors}

\begin{figure}[h]
\centering
\includegraphics[width=0.9\linewidth]{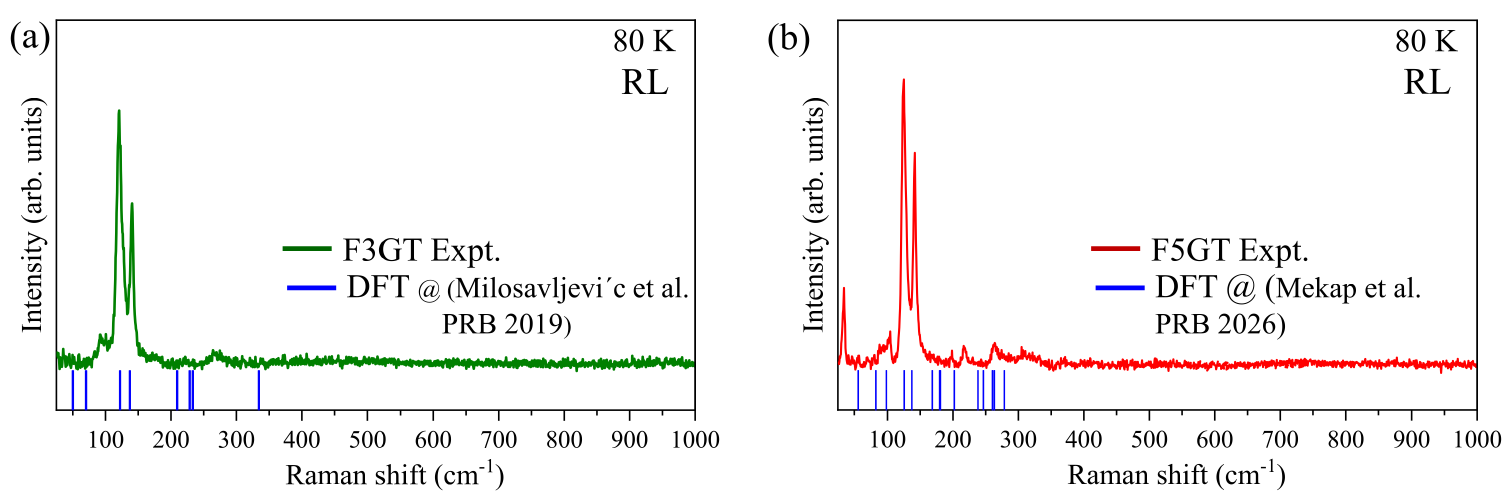}
\caption{ Background-subtracted Raman spectrum measured in RL polarized configuration  at $80$ K for (a) F3GT and (b) F5GT.
Theoretically obtained modes are marked by blue bars below. Difference between calculated and experimental phonon frequencies in F3GT/F5GT are expected, since in these report on first-principles calculations neglected finite-temperature lattice effects. More importantly, calculated phonon frequencies are sensitive on the assumed magnetic configuration; such discrepancies have been discussed for Raman-active modes of F3GT/F5GT in both the referred articles. } 
\label{fig:F5GT_DFT}
\end{figure}

\noindent We refer to DFT derived atomic displacement vector for these modes in\\ Ref: [58] , for TeO$_2$, \\
Ref: [59] 
for Te, \\Ref: [49]  for F3GT,\\ and Ref: [45] for F5GT. \\ 
The occurrence of Raman-active Te-related modes in the 120–145 cm$^{-1}$ range provides a natural explanation for why superficially similar spectral features can occur in chemically distinct Te-containing compounds. This is a consequence of the similar Te-Te bonding environments in tellurides, not evidence of surface contamination. Fig. \ref{fig:F5GT_DFT} compares the experimentally observed Raman spectrum in RL polarization configuration and the Raman frequency of the phonon modes from DFT calculations [Ref. F5GT: our work @ [45], F3GT: [49]]. Though due to the limitations of DFT to describe the F3GT and F5GT, the simulated peak positions do not match exactly; in both cases, one does not expect any spectral feature beyond 350 cm$^{-1}$, which we, too, find in our experimental spectra of these vdW compounds. 

\subsection{Raman specta of cleaved and uncleaved F3GT and F5GT}

\begin{figure}[h]
\centering
	\includegraphics[width=0.8\linewidth]{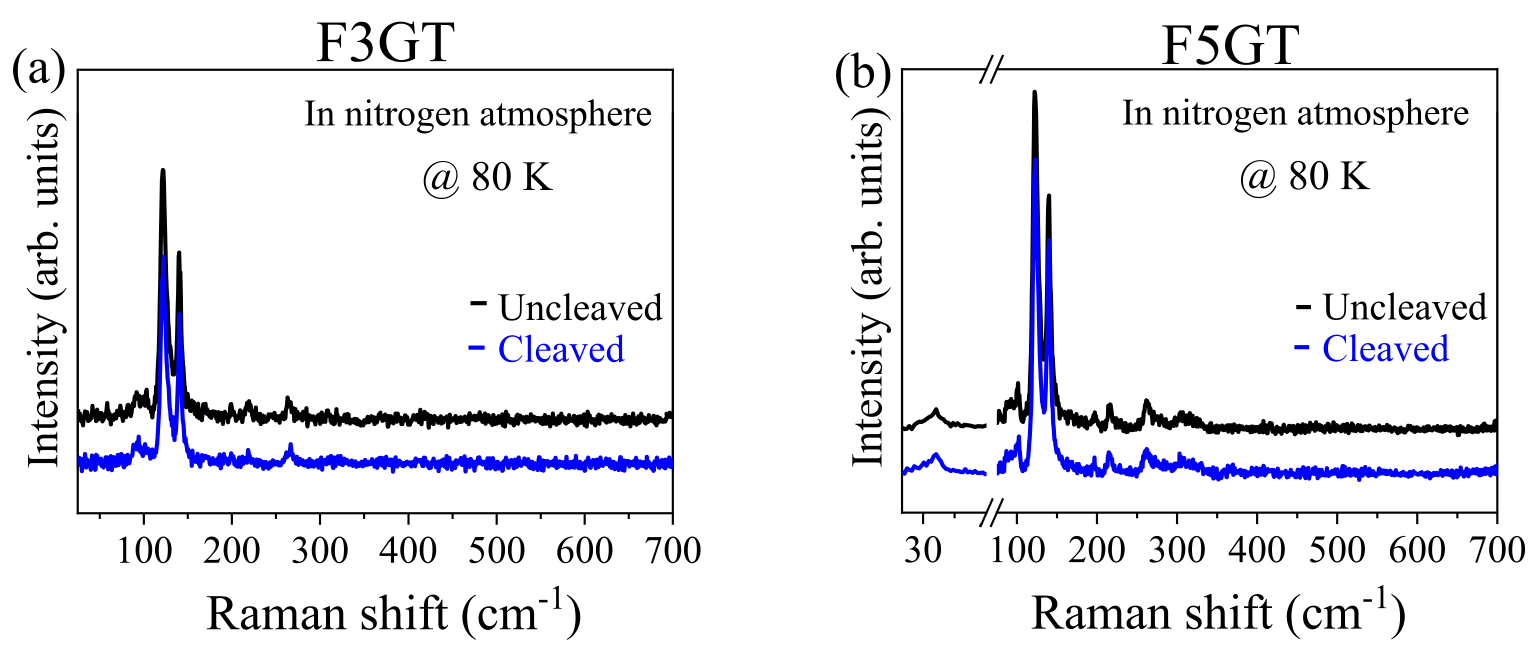}
	\caption{Unpolarized background-subtracted Raman spectra of cleaved and uncleaved  (a) F3GT and (b) F5GT, respectively,at 80 K.}
	
\label{fig:cleaved_FGT}
\end{figure}

\noindent Refer to Fig. \ref{fig:cleaved_FGT}. The reproducibility between freshly cleaved and uncleaved samples strongly suggests that the dominant contribution is not from a surface degradation product. The reproducibility of the spectra over the entire course of the study demonstrates that the observed modes are intrinsic to the bulk crystals.

\subsection{Wavelength-dependent Raman intensity of TeO$_2$, Te, F3GT, and F5GT}

\begin{figure}[h]
    \centering
    \includegraphics[width=0.75\linewidth]{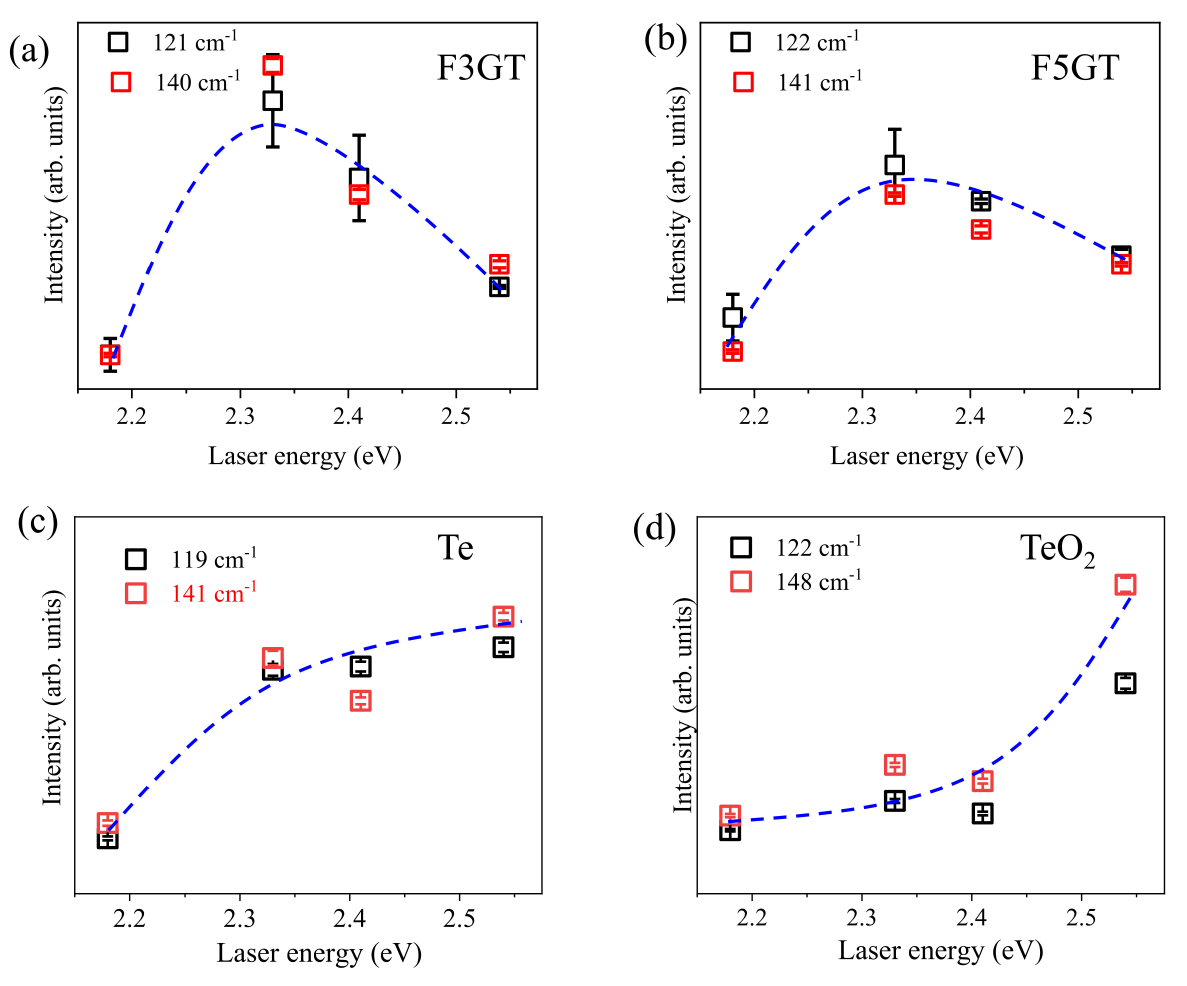}
    \caption{Evolution of the intensity of two intense peaks (over the range 110 to 170 cm$^{-1}$) with wavelength of (a) F3GT, (b) F5GT, (c) Te, and (d) TeO$_2$, as obtained from unpolarized Raman measurements.}
    \label{fig:wavelength_Te}
\end{figure}
\noindent Refer to Fig. \ref{fig:wavelength_Te} in which the wavelength-dependent intensities of two modes, under consideration, are plotted. We find a completely different trend in cases of F3GT/F5GT and TeO$_2$ and Te. Note that Te is a semimetallic compound, and experimental values of the band gap of TeO$_2$ are around 3.5 eV [60]. 

\subsection{Systematic differences in Raman spectral behavior of F3GT and F5GT}
\label{sec.systematic_diff}
The systematic variation of Raman spectra with composition and temperature, which we have reported in this article, confirms the intrinsic origin.
If the dominant peaks were from TeO$_2$ or elemental Te, we would expect:
- Identical spectral behavior for F3GT and F5GT, since the same degradation product would be present on both samples.
- No systematic temperature dependence related to magnetic transitions, since TeO$_2$ and Te are non-magnetic.
- No systematic variation with crystal structure or Fe content.

Contrary to this expectation, we observe:\\
- Systematic differences between F3GT and F5GT 
- Clear temperature-dependent anomalies at the magnetic transitions (Figs. 3, 7, and 8 in our manuscript). The Fano parameter ($1/q$), the chiral mode splitting ($\Delta f$), and the Raman asymmetry parameters all exhibit systematic changes at ($T_C$), demonstrating that the phonons are coupled to the magnetic order. A non-magnetic surface contaminant would not show such behavior.
- Systematic differences in polar plots (Figs. 4 and 5): F3GT exhibits four-lobed and elliptical polar patterns for E$^+$ and E$^-$ modes under resonance, while F5GT shows helicity-symmetric responses. These distinct behaviors arise from the different crystal structures and magnetic anisotropies of the two compounds, not from a common surface contaminant.

\subsection{Existence of chiral phonon modes}
\label{sec.chiral_phonon}
The "TeO$_2$ hypothesis" cannot explain the chiral phonon behavior.
Most importantly, the same hypothesis cannot explain the chiral phonon splitting ($\Delta f$) that we observe in both F3GT and F5GT (Fig. 3). The splitting of the doubly degenerate E modes into E$^+$ and E$^-$ components under left- and right-circularly polarized light is a direct consequence of time-reversal symmetry breaking in the magnetically ordered state. This is a symmetry-selective effect that depends on the magnetic point group of the crystal. TeO$_2$, being non-magnetic, would not exhibit a temperature-dependent splitting that tracks the magnetic order parameter. The fact that $\Delta f$ varies systematically with temperature and shows different behavior in F3GT (out-of-plane easy axis) versus F5GT (in-plane easy axis) provides evidence that the observed phonons are intrinsic to the magnetic crystals.

\subsection{Similar spectral profile for different laser power and different laser wavelength}
\begin{figure*}[h]
\centering
\includegraphics[width=0.5\linewidth]{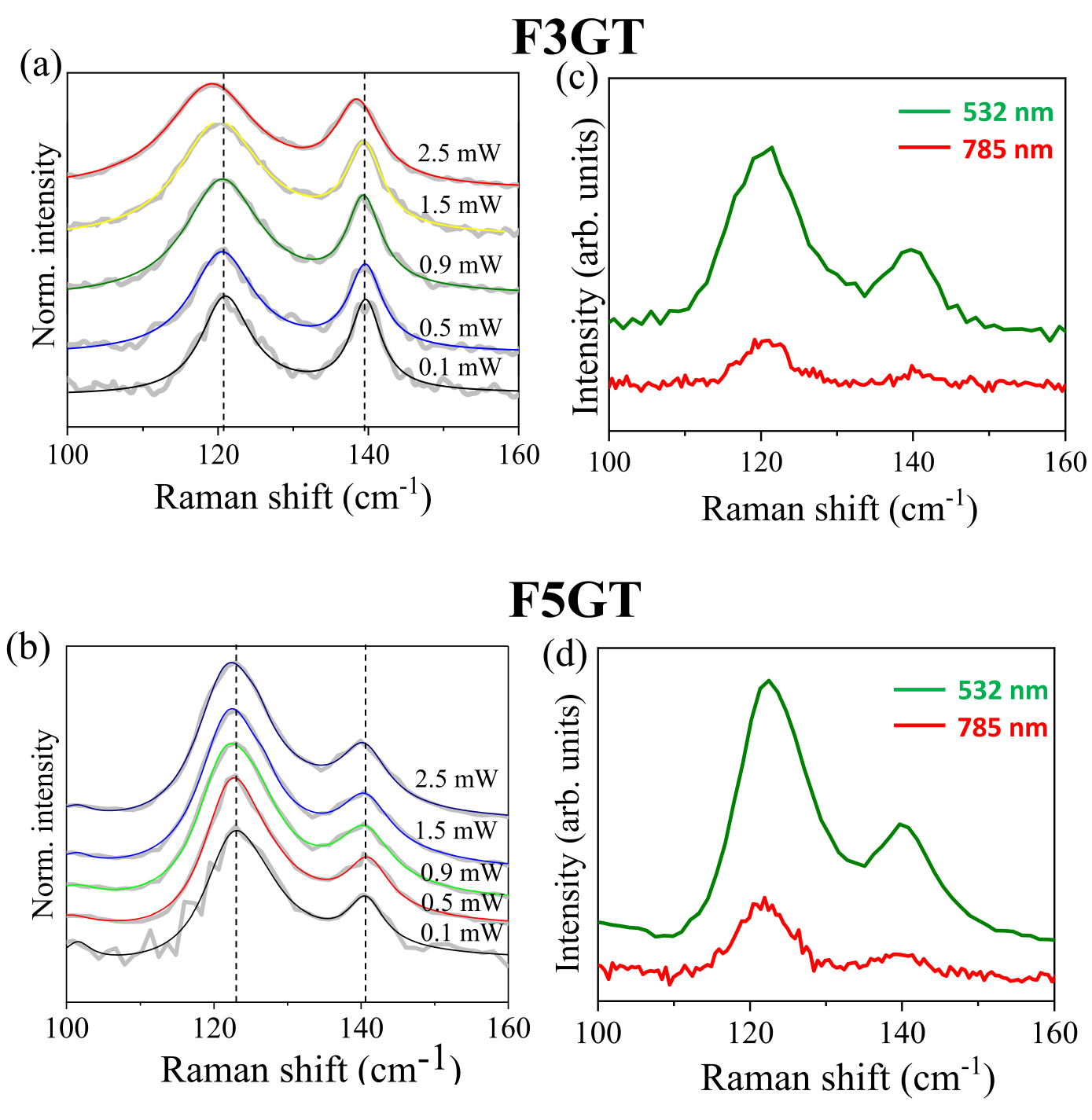}
	\caption{(Color online) Laser power-dependent unpolarized background-subtracted Raman spectra of (a) F3GT and (b) F5GT. Wavelength-dependent Raman spectra of (c) F3GT, (d) F5GT.}
	
\label{fig:power_wavelength}
\end{figure*}
The similarity of the spectral profiles across different experimental configurations demonstrates the reproducibility of the phonon spectra with respect to variations in laser power, excitation wavelength, and sample surface conditions. The surface degradation under laser power, or surface contamination, is expected to show differences with laser power and wavelength dependence, respectively. Refer to Fig. \ref{fig:power_wavelength}

\section{Raman background continuum}
\label{sec.background}
\subsection{Raman background continuum of F3GT and F5GT}
\begin{figure}[h]
    \centering
    \includegraphics[width=0.65\linewidth]{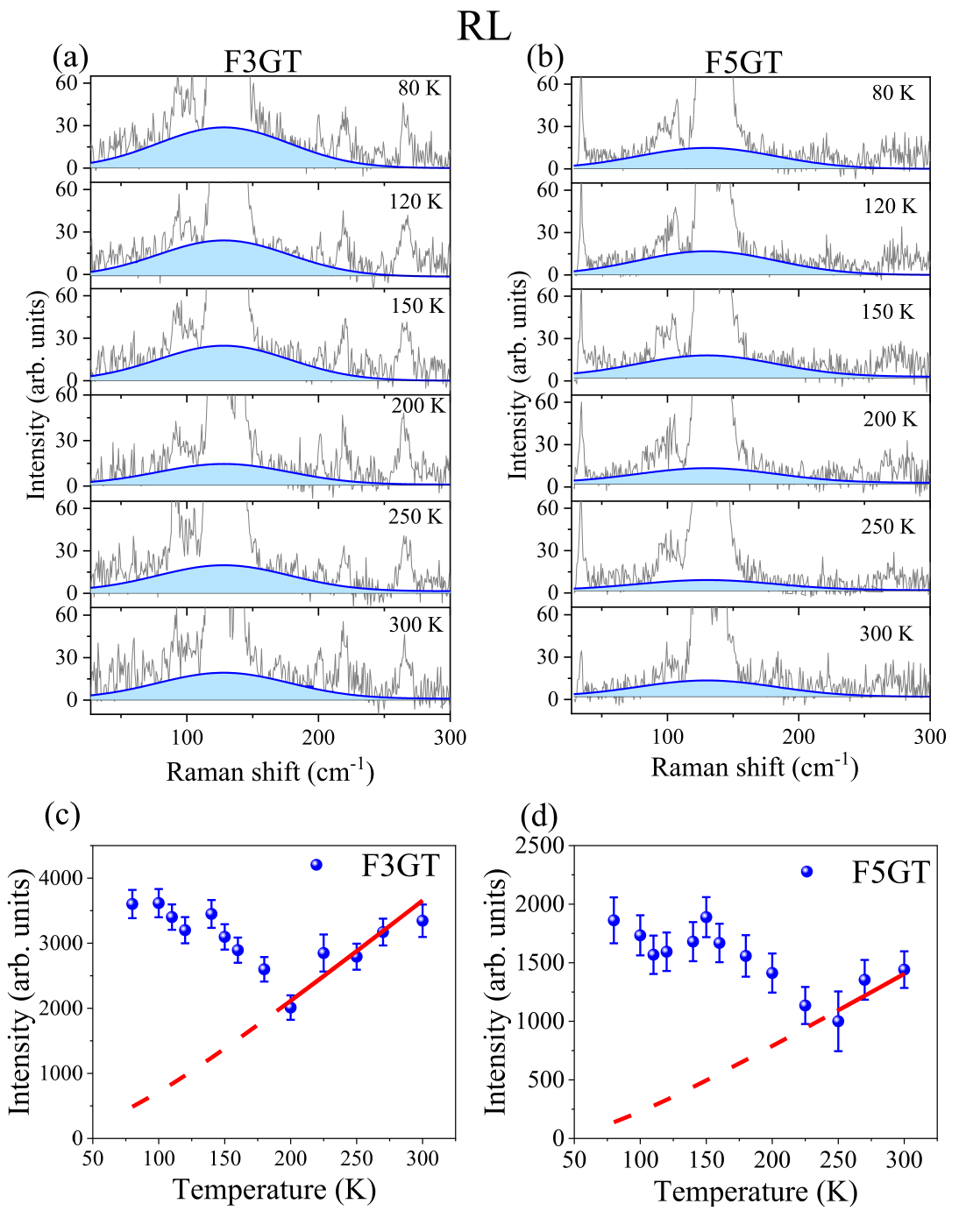}
    \caption{ Characteristics magnified Raman spectra in RL polarization configuration with a Gaussian background (blue shaded region) for (a) F3GT and (b) F5GT at different temperatures. The evolution of background intensity with temperature for (c) F3GT and (d) F5GT.  The red curves are the best fit to the data points using the Bose-Einstein thermal factor, which are then extrapolated down to 78 K, shown by dashed curves.}
    \label{fig:bg_spectra}
\end{figure}

 Figure \ref{fig:bg_spectra} (a,b) shows magnified Raman spectra of F3GT and F5GT over an extended frequency range in RL polarization configuration, where a broad temperature-dependent background is clearly visible in addition to the relatively sharp phonon features. The background extends well beyond the phonon frequencies of interest, up to approximately 300 cm$^{-1}$. In particular, the phonon modes near 125 cm$^{-1}$ and 142 cm$^{-1}$ in FGT are located within this broad continuum-like Raman background and can therefore, in principle, interfere with it.  A similar broad background is also observed in the LR polarization configuration.
It is to be noted that the observation of a broad continuum by itself does not prove that the phonon asymmetry must arise from Fano interference. It establishes, however, the essential experimental prerequisite for such an interaction: coupling of relatively sharp phonon excitation with a broad scattering continuum of overlapping energies. The Fano description provides a physically motivated way of describing the interference between these two scattering channels.

The temperature evolution of this background, Fig. \ref{fig:bg_spectra}(c,d) provides additional information. Its integrated intensity increases systematically with temperature in the high-temperature regime and can be described by a Bose-population-type temperature dependence (red solid curves). These lines are then extrapolated down to 78 K (red dashed curves). More importantly, the temperature-dependent excess background cannot be represented simply as a temperature-independent baseline extrapolated from the high-temperature spectrum. We regard this broad, temperature-dependent scattering contribution at low temperatures as consistent with a Raman continuum.

\newpage
\subsection{Wavelenth-depenent Raman background continuum of F3GT and F5GT}
\begin{figure}[h]
    \centering
    \includegraphics[width=0.8\linewidth]{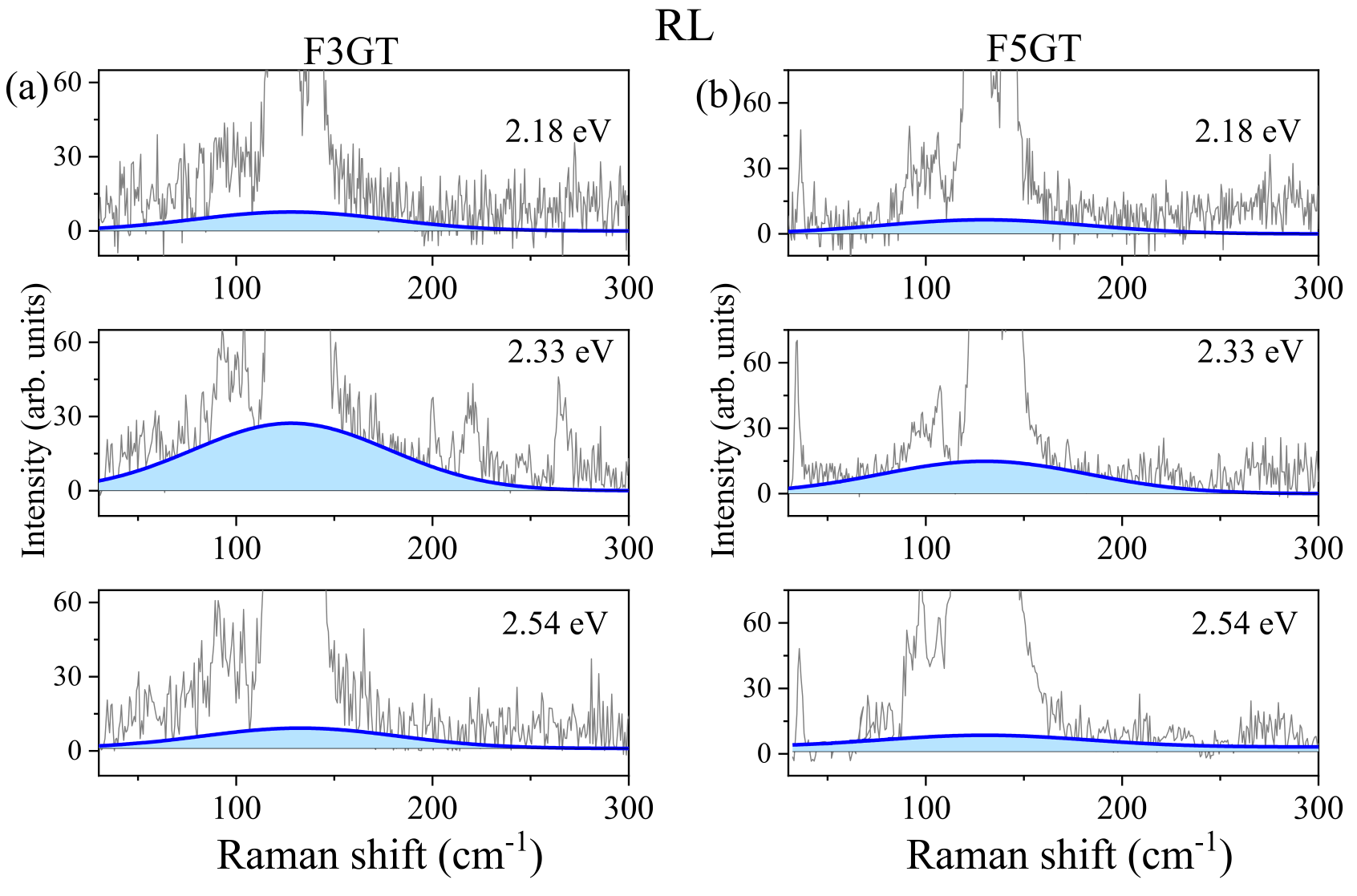}
    \caption{Magnified Raman spectra in RL polarization configuration  with a Gaussian background with three different laser excitation energies at 80 K for (a) F3GT and (b) F5GT }
    \label{fig:bg_wavelength}
\end{figure}

Fig. \ref{fig:bg_wavelength} (a,b) compares the spectral background recorded using two different excitation wavelengths. Indeed, we find that the background intensity drops in off-resonance conditions, consistent with an electronic origin. In this article, we have demonstrated, using wavelength-dependent Raman measurements, that all observed effects in the Raman spectral line shape arise from optical resonance. It is also worth noting that although the background intensity decreased under off-resonance conditions, we do not observe any shift in the profile, arguing against an interpretation of the broad background as photoluminescence.
\newpage
\subsection{Comparison of background continuum of FGT with TeO$_2$ }
\begin{figure}[h!]
    \centering
\includegraphics[width=0.35\linewidth]{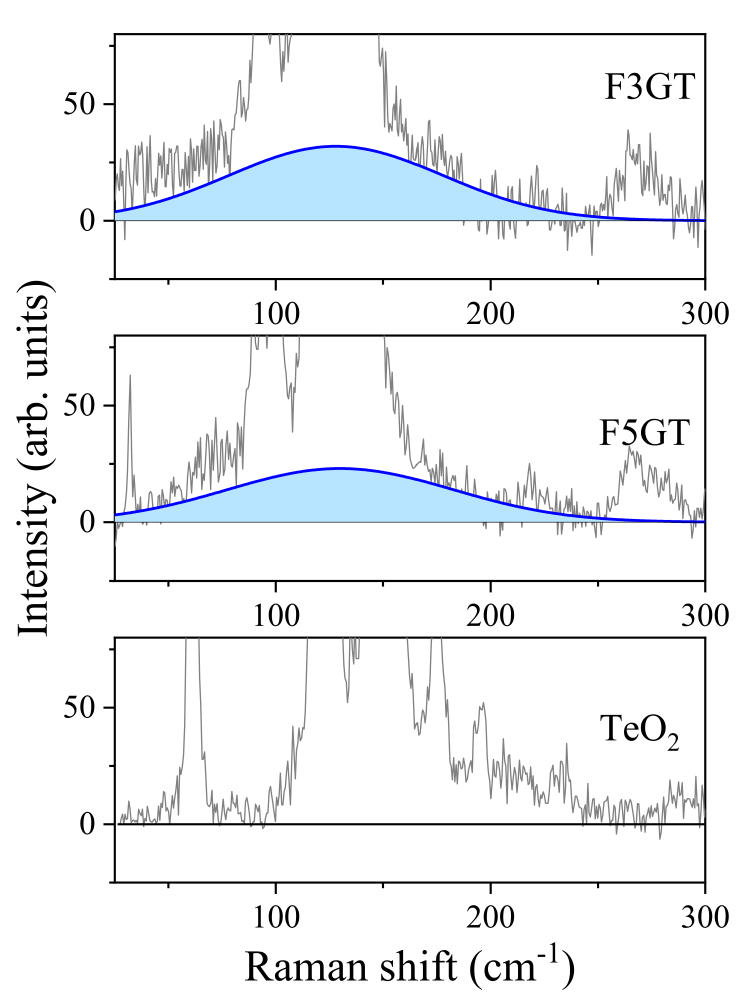}
    \caption{Comparison of the background in the unpolarized Raman spectra of F3GT, F5GT and TeO$_2$ at 300 K.}
    \label{fig:bg_TeO2}
\end{figure}
In Fig. \ref{fig:bg_TeO2} we compare the magnified Raman spectra of TeO$_2$ along with that of F3GT and F5GT. Spectra are recorded under identical experimental conditions. The flat background of TeO$_2$ proves that the background observed in Fig. \ref{fig:bg_spectra} is not the instrumental artifact.

\newpage
\subsection{Statistically consistent description for Fano function}
\label{sec.fano_fit}
\begin{figure*}[h]
\centering	\includegraphics[width=0.7\linewidth]{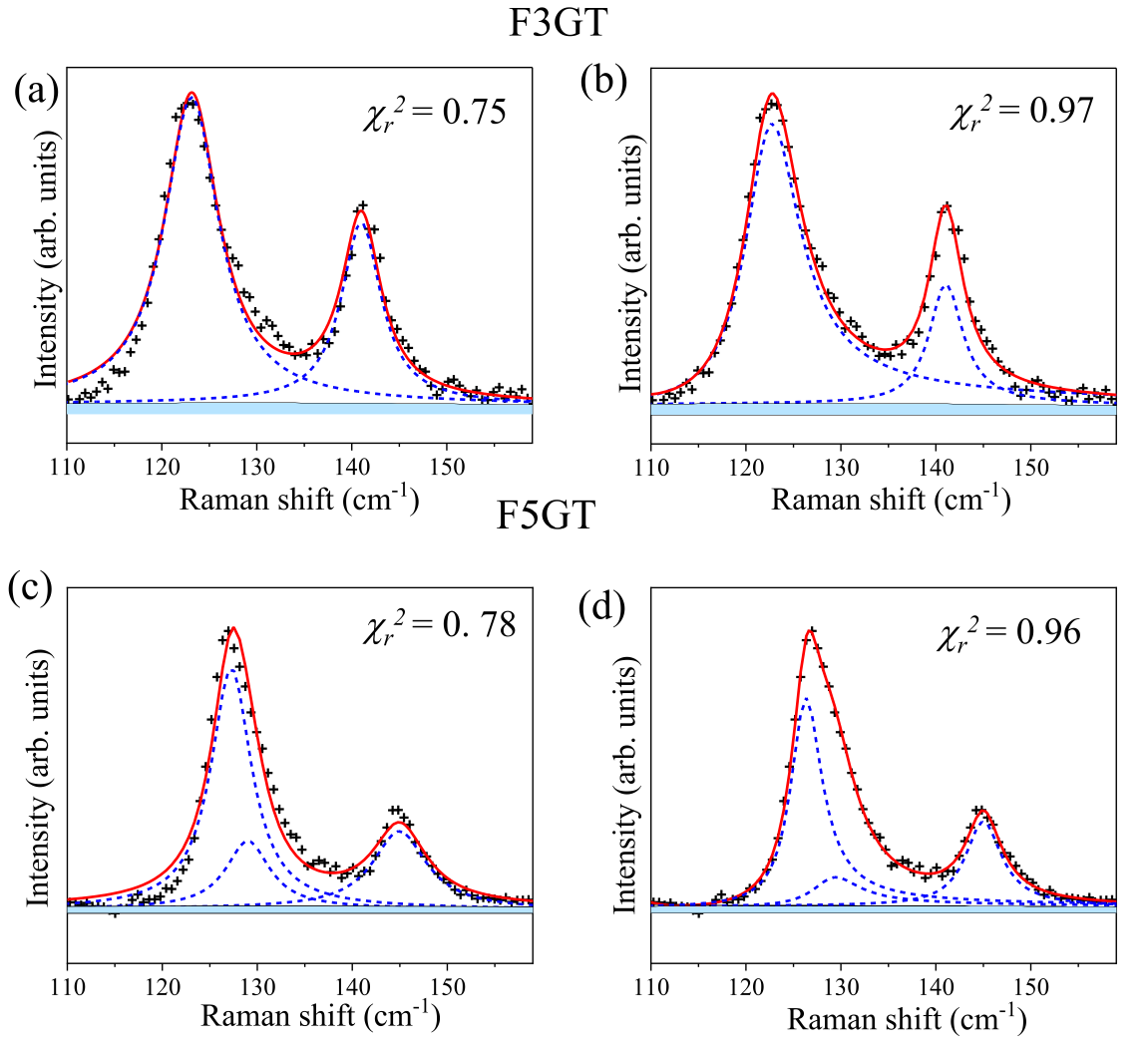}
	\caption{(Color online) Magnified view of Raman spectra at 80 K between 110 and 160 cm$^{-1}$ in RL polarization configurations. 
In all panels, ``+'' symbols are the spectral data. Red curves are the best fits to the data points using (a) a Lorentzian function for each of A$_1$, E modes, and (b) Fano line shape for F3GT; similarly for F5GT in (c) with three lorentzian function for  A$_1$(1),  A$_1$(2) and  E modes, and (d) with all Fano functions.
In all panels, the deconvoluted components are shown by the blue dashed curves. The values of $\chi^2$ as obtained from the best fit to the data points are available in the inset.}
	\label{Fano_fit}
\end{figure*}

We compared the Fano and Lorentzian descriptions quantitatively. For the representative spectra shown in Fig. \ref{Fano_fit}, the reduced $\chi^2$ obtained from the Fano fits is approximately 0.97, i.e. close to unity, indicating that the residuals are of the order expected from the experimental uncertainties. The corresponding reduced $\chi^2$ for the Lorentzian description is approximately 0.78.  The comparison demonstrates that the Fano function provides a statistically consistent description while allowing the experimentally observed asymmetric component to be quantified through  $1/q$ or justified by the existence of a broad background.

\clearpage
\newpage
\section{Symmetry assignments to Raman modes}
\label{sec.symmetry}

\begin{figure*}[h!]
\centering
	\includegraphics[width=0.5\linewidth]{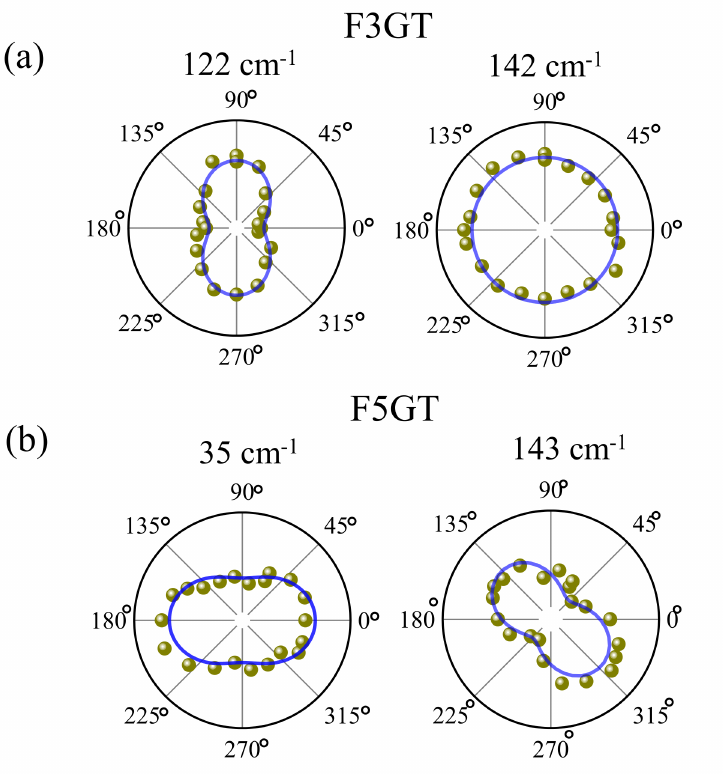}
	\caption{(Color online) Polar plot of normalized Raman intensity of observed peaks at (a) 122 cm$^{-1}$ and 142 cm$^{-1}$ of F3GT, and (b) 35 cm$^{-1}$ and 143 cm$^{-1}$ of F5GT under the vertical polarization of the incident light. Blue curves are the best fit to the data points using Eq.~\ref{F3Gt_int} for the E$_1$  mode and Eq. \ref{F3Gt_int_A1} for the A$_1$ mode of F3GT; and Eq.~\ref{F5Gt_E}, Eq.~\ref{eq.da1mod} for the E(1) and A$_1$(2) modes of F5GT.} 
	\label{fig:linear_symmetry}
\end{figure*}

According to the Placzek theory, the Raman intensity is given by $I \propto \left| \hat{e}_{s}^{T} \cdot \Re \cdot \hat{e}_{i} \right|^2$, where $\hat{e}_i$ and $\hat{e}_s$ are the polarization unit vectors of the incident and scattered light, respectively.
For our experimental configuration, the crystal basis for the vertically polarized incident light are
$X = ( 1 \; 0 \; 0 )^{T} \quad \text{and} \quad Y = ( 0 \; 1 \; 0 )^{T} $.
For linearly polarized light, the angle($\varphi$)-dependent scattered intensity (for $\hat{e}_{i} = (0 \;1 \;0)^{T}$ and $\hat{e}_{s}^{T} =(\cos\varphi \;\sin\varphi \;\;0)$.

Considering the  $6/mmm$ magnetic point group for the $P6_{3}/mmc$ space group for F3GT, the Raman tensors for A$_1$, and the degenerate E modes can be written as: [69] 

\begin{eqnarray}
\label{Eqn_6m}
\nonumber
\Re(\text{A}_{1}) &=& \left( \begin{array}{ccc}
    a & 0 & 0 \\
     0 & a & 0 \\
     0 & 0 & b \\
    \end{array}\right), \quad
   \quad
   \end{eqnarray}
Doubly degenerate Raman tensors of E$_{2g}$ mode are
\begin{eqnarray}
\Re(\text{E}{_{2g}^{1}}) &=&
\scalebox{0.85}{$
\left(
\begin{array}{ccc}
 c & 0 & 0 \\
0 & -c & 0 \\
 0 & 0 & 0
\end{array}
\right)
$},
\quad
\Re(\text{E}_{2g}^{2}) =
\scalebox{0.85}{$
\left(
\begin{array}{ccc}
0 & c & 0 \\
 c & 0 & 0 \\
 0 & 0 & 0
\end{array}
\right)
$}.
\end{eqnarray}
In the ferromagnetic phase with broken TRS, the dielectric susceptibility is no longer symmetric,
\begin{equation}
\nonumber
\chi_{ij}=\chi_{ij}^{S}+\chi_{ij}^{A}
\end{equation}
Since the antisymmetric component is odd under time reversal symmetry
\begin{equation}
\nonumber
\chi_{ij}^{A}\propto \epsilon_{ijk}M_k
\end{equation}
and can be considered as the leading magnetization-odd contribution allowed by the symmetry.
For $M||z$
\begin{eqnarray}
\chi_{ij}^{A} &=&
\scalebox{0.85}{$
\left(
\begin{array}{ccc}
0 & ig & 0 \\
-ig & 0 & 0 \\
 0 & 0 & 0
\end{array}
\right)
$},
\quad
\end{eqnarray}
The general form of Raman tensors becomes
\begin{equation}
\nonumber
R(E^{+})=R_{0}(E^{+})+R_{M}(E^{+})
\end{equation}
\begin{equation}
 \nonumber   
R(E^{-})=R_{0}(E^{-})+R_{M}(E^{-}).
\end{equation} 
Thus,
\begin{equation}
\begin{aligned}
\Re (\mathrm{E}^{+}) &=
\scalebox{0.85}{$
\left(
\begin{array}{ccc}
 c & id & 0 \\
 -id & -c & 0 \\
 0 & 0 & 0
\end{array}
\right)
$}
\qquad
\Re(\mathrm{E}^{-}) =
\scalebox{0.85}{$
\left(
\begin{array}{ccc}
 id & c & 0 \\
 c & -id & 0 \\
 0 & 0 & 0
\end{array}
\right)
$}.
\end{aligned}
\label{Eqn_6m_e+}
\end{equation}

Hence, the Raman scattering cross-section in the lab frame can be calculated for the A$_1$ modes and E modes as, 

\begin{eqnarray}
I ^{L}(\Re(\text{A}_{1}) , \varphi ) = \vert a \sin \varphi \vert^{2}, \quad \text{and} \quad I^{L} ( \Re(\text{E$^+$}), \varphi ) = \vert d \vert^{2}, I^{L} (\Re (\text{E$^-$}), \varphi ) =  \vert c \vert^{2} .
\label{F3Gt_int}
\end{eqnarray}
Thus, from our polar plots in Fig.~\ref{fig:linear_symmetry} (a), the Raman peak at 142 cm$^{-1}$ can be ascribed to the symmetry E$_1$. The polar plot of the Raman peak at 122 cm$^{-1}$ is nearly bi-lobed  following the Raman tensor 

\begin{eqnarray}
\label{Eqn_6m_a1}
\nonumber
\Re(\text{A}_{1}) &=& \left( \begin{array}{ccc}
    a & if & 0 \\
     -if & a & 0 \\
     0 & 0 & c \\
    \end{array}\right), \quad
   \quad
\end{eqnarray}
considering the antisymmetric component $\chi_{ij}^{A}$.
The modified intensity of the A$_1$ mode is given by
\begin{eqnarray}
I ^{L}( \Re(\text{A}_{1}) , \varphi ) = \vert a \sin \varphi \vert^{2}+ \vert f\cos\varphi\vert^2
\label{F3Gt_int_A1}
\end{eqnarray}

\vspace{0.5cm}

For F5GT, for considering 3$m$ magnetic point group the tensor can be written as [69]:
\begin{eqnarray}
\Re (A_{1}) = \left( \begin{array}{ccc}
    a & if & 0 \\
     -if & a & 0 \\
     0 & 0 & b \\
    \end{array}\right), \quad \text{and} \quad 
\Re (E^{+}) = \left( \begin{array}{ccc}
    d & id & ic \\
     id & -d & c \\
     ig & g & 0 \\
    \end{array}\right) 
   \Re (E^{-}) = \left( \begin{array}{ccc}
    e & -ie & -ij \\
     -ie & -e & j \\
     -ih & h & 0 \\
    \end{array}\right) .
\end{eqnarray}
For linearly polarized light, the angle dependent intensity can be found as
\begin{eqnarray}
\label{eq.da1}
I^{L} (\Re(\text{A}_{1}) , \varphi ) = \vert a \sin \varphi \vert^{2}+ \vert f \cos \varphi \vert^{2}
\end{eqnarray}
and
\begin{eqnarray}
I^{L} (\Re(\text{E}^{+}),\varphi ) = \vert d\vert^{2}, \quad \text{and} \quad I^{L} (\Re(\text{E}^{-}),\varphi ) =\vert e\vert^{2}.
\end{eqnarray}

For the incoherent superposition of two degenerate E modes\\
\begin{eqnarray}
\label{eq.dephi}
I^{L}(\Re(\text{E}),\varphi)= \vert I^{L} (\Re(\text{E}^{+}),\varphi ) \vert+\vert I^{L}(\Re(\text{E}^{-}),\varphi)\vert= \vert d \vert^{2} +\vert e \vert^{2},
\end{eqnarray}

The elliptical polar plot of the intensity of the peak at 35 cm$^{-1}$ could be explained by considering the coherent superposition of two degenerate E modes under resonance conditions [45]. The tilt in the nearly bi-lobed polar plot Raman of the Raman intensity arises from modifications to the Raman tensor introduced due to electron–phonon coupling for the A$_1$ mode [45].

For E mode, considering the coherent superposition of two E modes [45]:
\begin{eqnarray}
\label{eq.dephi2} I^{L}(\Re(\text{E}),\varphi)= \vert u.(\Re\text{E}^{+})+v.(\Re\text{E}^{-} ) \vert^{2} =\vert u.d\vert^{2}+\vert v.e\vert^{2}+2\vert u.d\vert \vert v.e\vert\sin^{2} \varphi-2\vert u.d\vert \vert v.e \vert \cos^{2}\varphi.
\label{F5Gt_E}
\end{eqnarray}

For A$_1$ mode [45]
\begin{eqnarray}
\nonumber \Re(A_1^{mod}) = \left( \begin{array}{ccc}
    a \exp \left( {i\alpha_{a}} \right) & if & 0 \\
     -if & a \exp \left( {i\alpha_{a^{}}} \right) & 0 \\
    0 & 0 & b \exp \left( {i\alpha_{b}} \right) \\
  \end{array}\right)
\end{eqnarray}

\begin{eqnarray}
\label{eq.da1mod}
I^{L} ( \Re (A_1^{mod}), \varphi ) = \vert a \sin \varphi*\cos\alpha_{a}\vert^{2}+ \vert a\sin \varphi* \sin \alpha_{a}+f \cos \varphi \vert^{2}
\end{eqnarray}

\clearpage
\newpage
\section{Temperature-dependent Raman studies under RL and LR configurations}
\label{sec.tempr_depnd}

\subsection{Temperature-dependent Raman shift in F3GT and F5GT under RL and LR polarization configurations}
\begin{figure}[h]
    \centering
    \includegraphics[width=0.7\linewidth]{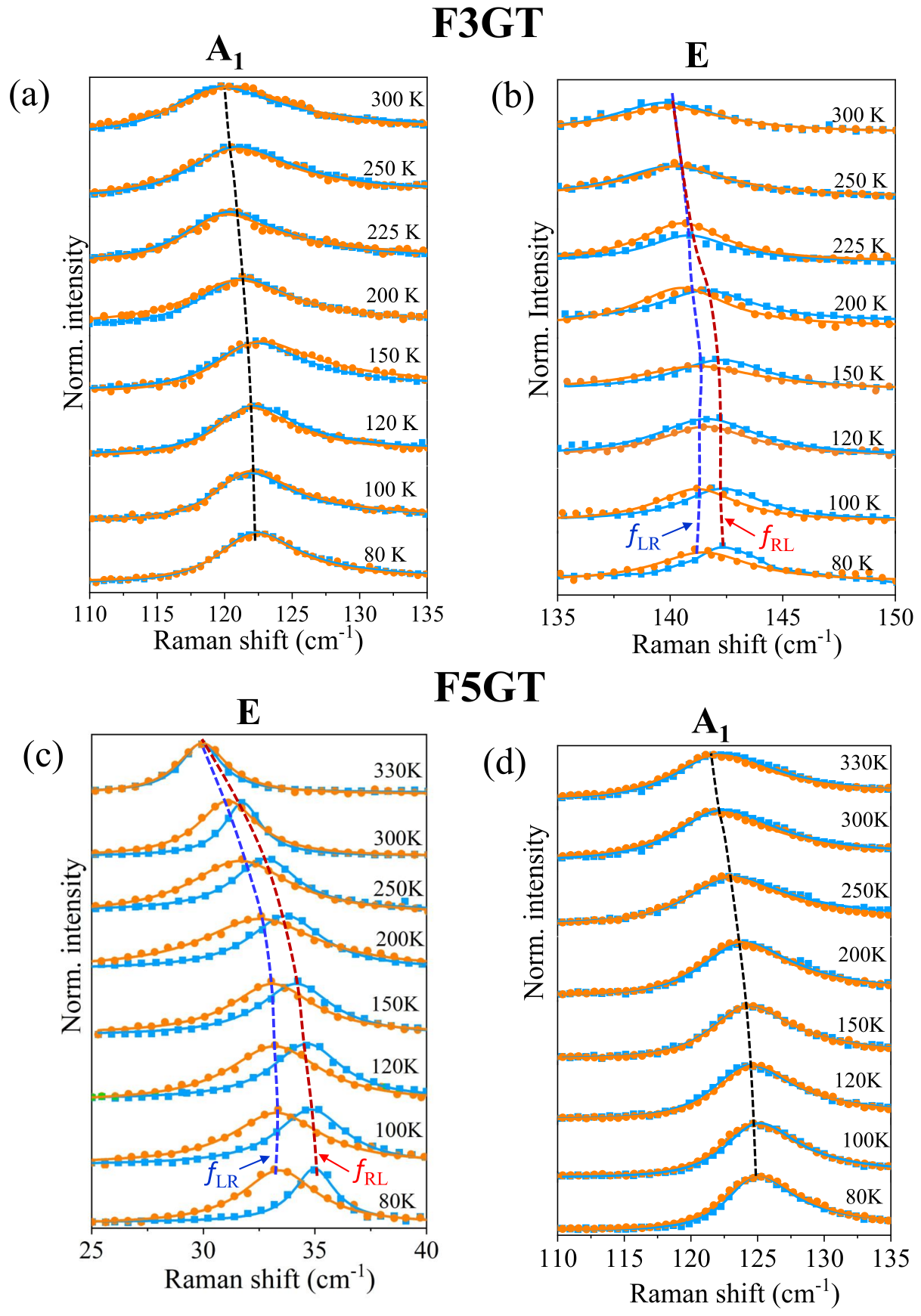}
    \caption{Helicity-resolved Raman spectra
of the (a) A$_1$  and (b) E mode for various temperatures for F3GT, same for F5GT in (c) and (d), respectively, after subtracting the fitted background. Yellow symbols and curves correspond to spectra recorded in LR configuration. The same in RL polarization configurations are shown by blue symbols and curves. The red and blue dashed curves mark the temperature-dependent shifts in the spectra for the RL and LR configurations.
 }
    \label{fig:splitting_fgt}
\end{figure}

\newpage
\section{Wavelength- dependent Raman measurements}
\label{sec.wavelength_depnd}
\subsection{Wavelength-dependent polarized experiment}
\begin{figure*}[h]
\centering
	\includegraphics[width=0.8\linewidth]{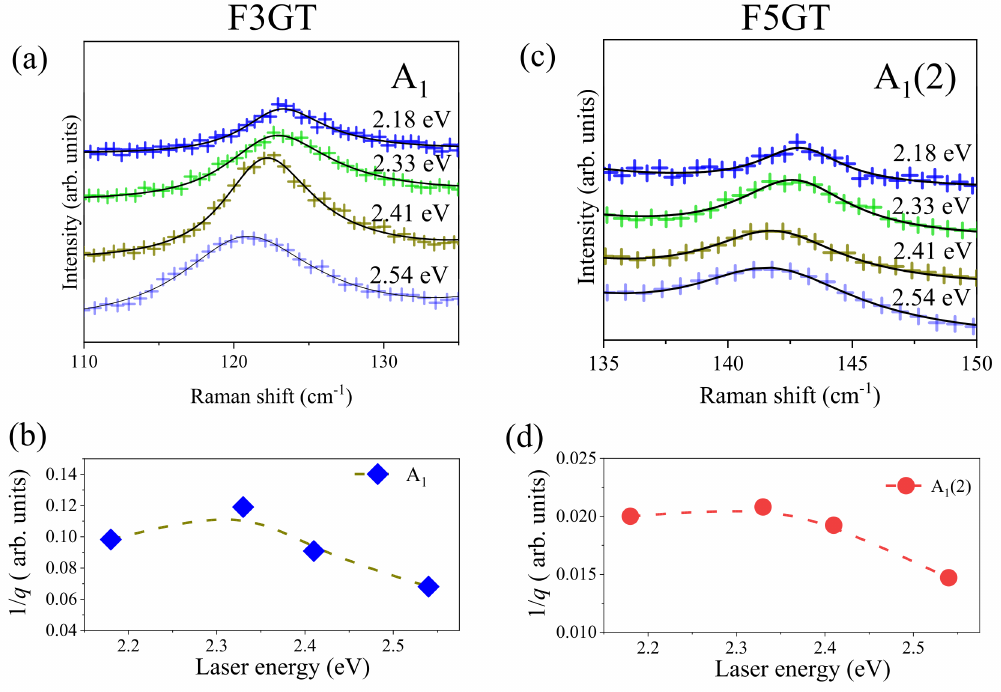}[h]
	\caption{(Color online)  Spectra recorded at different laser excitation energies, marked on the right, in RL configuration for (a) F3GT, and  (c) F5GT. Variation of 1/$q$ for the A$_1$ modes with different laser excitation energies for (b) F3GT, and (d) F5GT.}
	\label{fig:wavelength_polar}
\end{figure*}

\section{Fitted parameters for F5GT in Fig. 5 using Eqn. 5 of the Main text}
\label{sec.fitted_parameter}

\begin{table}[h]
\centering
\caption{Measured values under different temperatures and polarization states.}
\label{tab:alpha_values}
\begin{tabular}{|c|c|c|c|c|}
\hline
 A$_1$(2)& \textbf{80 K} & \textbf{80 K} & \textbf{300 K} & \textbf{300 K} \\
\cline{2-5}
 & \textbf{LCP} & \textbf{RCP} & \textbf{LCP} & \textbf{RCP} \\
\hline
$\alpha$ (rad) & 0.74& 0.95& 1.06& 1.39\\
\hline
$\alpha_{a\prime a}$ (rad) &2.16 & 4.18& 2.06& 2.44\\
\hline
\end{tabular}
\label{fitted_parameter}
\end{table}
The above parameters are effective phases within the phenomenological polarization model used by us, in Eqn. 5 of the main text, rather than direct crystallographic angles. These parameters only provide the relative phase between different Raman tensor contributions required to reproduce the observed angular intensity pattern.

\end{document}